\documentclass[%
 reprint,
superscriptaddress,
nofootinbib,
 amsmath,amssymb,
 aps,
pra,
]{revtex4-2}
\usepackage{lipsum} 
\usepackage{comment,color,xcolor}
\usepackage{graphicx}
\usepackage{dcolumn}
\usepackage{bm}

\graphicspath{{./},{./figures/}}

\begin{document}

\preprint{APS/123-QED}

\title{Particle approximations of Wigner distributions for $n$ arbitrary observables}

\author{Ralph Sabbagh}
\author{Olga Movilla Miangolarra} 
\affiliation{Department of Mechanical and Aerospace Engineering, University of California, Irvine, California 92697, USA}
\author{Hamid Hezari}
\affiliation{Department of Mathematics, University of California, Irvine, California 92697, USA}
\author{Tryphon T. Georgiou}
\affiliation{Department of Mechanical and Aerospace Engineering, University of California, Irvine, California 92697, USA}

\begin{abstract}
A class of signed joint probability measures for $n$ arbitrary quantum observables is derived and
studied based on quasi-characteristic functions with symmetrized operator orderings of Margenau-Hill type. It is shown that the Wigner distribution associated with these observables can be rigorously approximated by such measures. These measures are given by affine combinations of Dirac delta distributions supported over the finite spectral range of the quantum observables and give the correct probability marginals when coarse-grained along any principal axis.  We specialize to bivariate quasi-probability distributions for the spin measurements of spin$-1/2$ particles and derive their closed-form expressions. As a side result, we point out a connection between the convergence of these particle approximations and the Mehler-Heine theorem. Finally, we interpret the supports of these quasi-probability distributions in terms of repeated thought experiments.
\end{abstract}

\maketitle


\section{\label{sec:1}Introduction}

In $1986$, L. Cohen and M. O. Scully developed bivariate quasi-probability distributions for the spin measurements of spin$-1/2$ particles \cite{Cohen1986-COHJWD}. Therein, two quasi-probability distributions based on quasi-characteristic functions induced from symmetrized operator orderings were studied. The first, termed the Margenau-Hill distribution, is given by the Fourier transform of the quasi-characteristic function 
\begin{align*}
f_{\text{MH}_1}(\xi_1,\xi_2) = \text{tr}\left(\rho \cfrac{e^{i\xi_1\hat{S}_1}e^{i\xi_2\hat{S}_2}+e^{i\xi_2\hat{S}_2}e^{i\xi_1\hat{S}_1}}{2}\right),
\end{align*}
where $\rho$ is a density matrix and $\hat{S}_1$ and $\hat{S}_2$ are the spin operators along two arbitrary directions in the Bloch sphere. The second, termed the Wigner distribution, is given by the Fourier transform of the quasi-characteristic function 
\begin{align*}
f_{\text{W}}(\xi_1,\xi_2) = \text{tr}\left(\rho e^{i\xi_1\hat{S}_1+i\xi_2\hat{S}_2}\right),
\end{align*}
and is the analog of the standard Wigner distribution for spin observables $\hat{S}_1$ and $\hat{S}_2$.
In $1992$, C. Chandler et al. derived the trivariate counterparts, with spin observables along mutually orthogonal directions \cite{chandler1992quasi}. Therein, it was shown that the computation for the trivariate Wigner distribution is simpler than its bivariate counterpart. 

In 2020, R. Schwonnek and R. F. Werner studied the Wigner distribution for an arbitrary tuple of bounded Hermitian operators $(\hat{A}_1,\hdots,\hat{A}_n)$ on a finite-dimensional Hilbert space \cite{10.1063/1.5140632}, and defined it as the Fourier transform of the quasi-characteristic function
\begin{align*}
f_{\text{W}}(\xi) = \text{tr}\left(\rho e^{i\xi\cdot \hat{A}}\right),
\end{align*}
where $\xi\cdot\hat{A}=\sum_{k=0}^n\xi_k\hat{A}_k,~\xi\in\mathbb{R}^n$. The distribution is termed ``Wigner" because it recovers the standard Wigner distribution when specialized to the canonical pair $\hat{A}_1 = \hat{X}$ and $\hat{A}_2 = \hat{P}$. Many of its basic properties, such as the support, location of singularities, positivity, and behavior under symmetry groups, were studied and illustrated with examples.
\begin{figure}
    \centering
    \includegraphics[width=\linewidth]{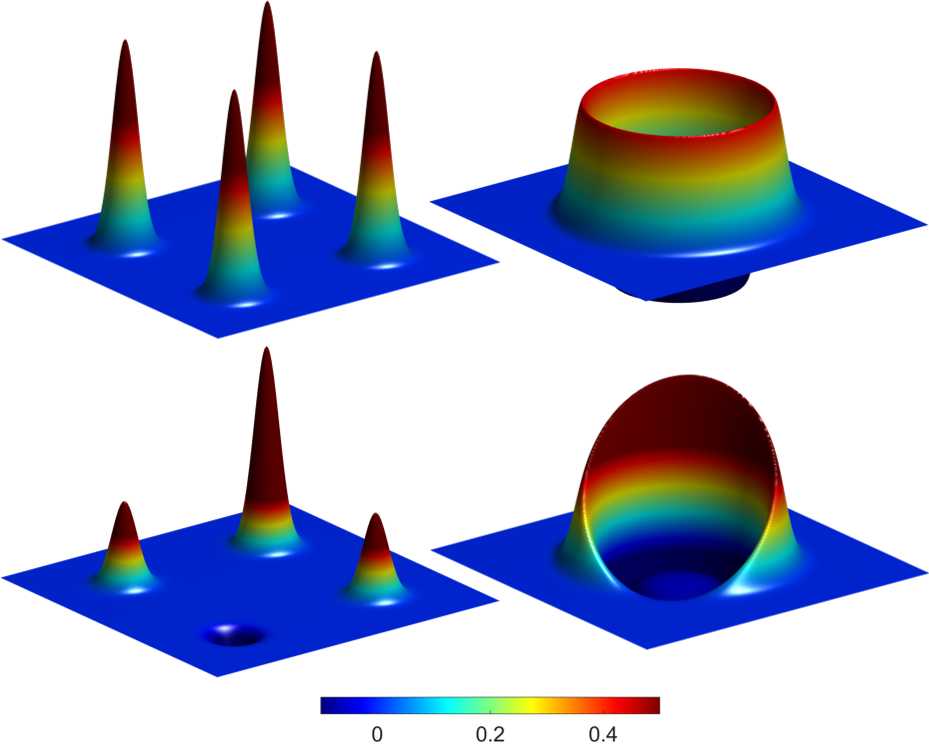}
    \caption{Gaussian-regularized $(\varepsilon = 0.01)$ bivariate quasi-probability distributions for the spin$-1/2$ observables along the $x$ and $y$ directions, at the state $\rho = \hat{I}/2$ (top) and $\rho = 0.5[1~(1-i)/\sqrt{2};~(1+i)/\sqrt{2}~1]$ (bottom) represented in the $\hat{S}_z$ eigenbasis. Left: Margenau-Hill quasi-probability distribution $p_{\text{MH}_1}$ consisting of $4$ Dirac delta distributions supported over the cartesian product of spin$-1/2$ eigenvalues $(\pm\hbar/2)\times(\pm\hbar/2)$. Right: Wigner quasi-probability distribution $p_{\text{W}}$ supported on a Disk of radius $\hbar/2$ with a complicated singularity near the boundary.}
    \label{fig:CohenScully}
\end{figure}

In many aspects, the standard Wigner distribution, defined by
\begin{align*}
p_{\text{W}} := \frac{1}{(2\pi)^n}\mathcal{F}(f_{\text{W}}),
\end{align*}
where $\mathcal{F}(\cdot)$ is the Fourier transform, has lent itself as a convenient choice for the phase-space representation of quantum states. This, in large part, is due to the Fourier duality of the canonical pair and their continuous spectra. It is often represented by a bounded and continuous function that integrates to one and is sign-indefinite: a salient non-classical feature. In contrast, when the same definition is applied to an $n$-tuple of non-commuting matrices such as in Refs. \cite{Cohen1986-COHJWD,chandler1992quasi,10.1063/1.5140632}, the non-classicality becomes three-fold. Not only is the distribution sign-indefinite, but it is no longer a measure to begin with, and the support need not be discrete. 
 \begin{figure}[t!]
    \centering
    \includegraphics[width=\linewidth]{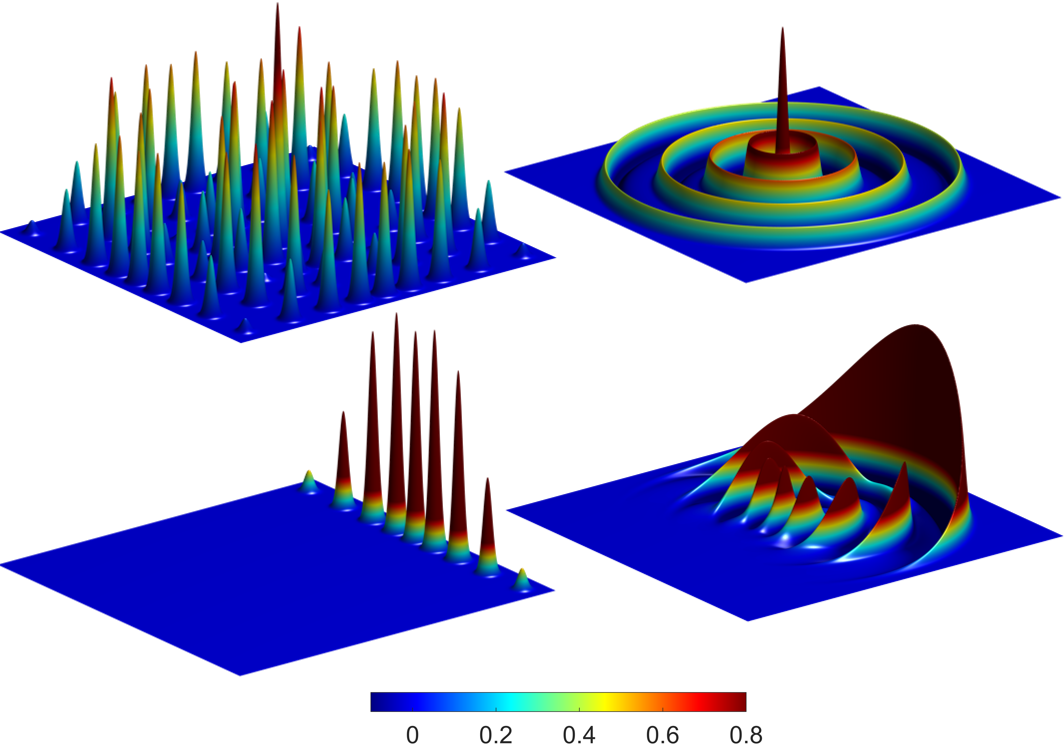}
    \caption{Gaussian-regularized $(\varepsilon = 0.1)$ bivariate quasi-probability distributions for the spin$-4$ observables along the $x$ and $y$ directions, at the maximally mixed state (top) and the $+4\hbar$ eigenstate along the $y$ direction (bottom). Left: Margenau-Hill quasi-probability distribution $p_{\text{MH}_1}$ consisting of $81$ Dirac delta distributions, each of which is supported over a cartesian product of spin eigenvalues, namely $(\pm k_1\hbar)\times(\pm k_2\hbar)$ where $k_1,k_2\in\{0,\hdots,4\}$. Right: Wigner quasi-probability distribution $p_{\text{W}}$ supported on a Disk of radius $4\hbar$ with singularities near concentric rings with radii $k\hbar,~k=0,\hdots,4$.}
    \label{fig:CohenScully2}
\end{figure}
The last two features are non-classical because the distribution is intended to be a joint law on the outcomes of simultaneously measuring the discrete observables. Indeed, when the observables do commute, $p_{\text{W}}$ is a classical discrete law. In general, however, it is a distribution supported beyond its intended set, with a rich singularity structure that is intimately related to the eigenvalues of the associated observables \cite{10.1063/1.5140632}. For instance, while the Margenau-Hill distribution studied in Ref. \cite{Cohen1986-COHJWD} is a discrete measure on the set of spin measurement outcomes $(\pm\hbar/2,\pm\hbar/2)$, the Wigner distribution for the same operators is supported on a Disk of radius $\hbar/2$, with a complicated singularity near the boundary. Gaussian-regularized\footnote{See the end of Sec. \ref{Sec:III} for details.} plots for both distributions are shown in Fig. \ref{fig:CohenScully} and analogs for a spin$-4$ particle are shown in Fig. \ref{fig:CohenScully2}.

Although the Wigner distribution $p_{\text{W}}$ lacks basic classical features, i.e., being a measure and having discrete support, its most remarkable classical feature still stands. It is the unique joint distribution for which the marginals of all linear combinations of the observables coincide with their quantum counterparts \cite{10.1063/1.5140632}. In contrast, the Margenau-Hill distribution in Refs. \cite{Cohen1986-COHJWD,chandler1992quasi} is a discrete measure that is supported over the classical set of measurement outcomes of the observables, but does not give the correct probability marginals for all linear combinations like $p_{\text{W}}$ does. Thus, each distribution possesses classical features expected from a joint probability distribution as well as non-classical features arising from the non-commutativity of the observables. A study of these features began in Ref. \cite{10.1063/1.5140632} for the Wigner distribution $p_{\text{W}}$, and in this work, we examine the features of a class of Margenau-Hill counterparts denoted by  $p_{\text{MH}_m}$, where $m\in\mathbb{N}$.

Specifically, we introduce, analyze, and interpret the quasi-probability distributions
\begin{align*}
p_{\text{MH}_{m}} = \cfrac{1}{(2\pi)^n}\mathcal{F}(f_{\text{MH}_m}),~m\in\mathbb{N},
\end{align*}
where  $f_{\text{MH}_m}$ are quasi-characteristic functions defined 
for an arbitrary tuple of hermitian matrices $(\hat{A}_1,\hdots,\hat{A}_n)$ and a quantum state $\rho$, see Sec.~\ref{sec:MH}, Eq.~\eqref{eq:MH}. It is shown that these distributions are real-valued, signed, and discrete probability measures given by affine combinations of Dirac delta distributions, and give the correct probability marginal when coarse-grained along any principal axis.

We shall refer to $p_{\text{MH}_m}$ as the Margenau-Hill quasi-probability distribution of order $m$, and show that
\begin{align*}
\lim_{m\rightarrow\infty}p_{\text{MH}_m} =p_{\text{W}},
\end{align*} in a suitable topology that can be upgraded to that of uniform convergence if the distributions are smeared with an appropriate Schwarz function. And so, while the general Wigner distribution $p_{\text{W}}$ is not a quasi-probability measure for the associated observables, it is not far from being one. 

Lastly, we specialize to pairs of spin$-1/2$ observables along orthogonal directions and derive closed-form expressions for $p_{\text{MH}_m}$ for any $m\in\mathbb{N}$. As a side result, we point out therein a connection between the convergence of the particle approximations $p_{\text{MH}_m}$ to $p_{\text{W}}$ and the Mehler-Heine theorem. We conclude by proposing an interpretation for the supports of the distributions $p_{\text{MH}_m}$ and $p_{\text{W}}$ in terms of repeated experiments.

\section{Preliminaries}
 In this section, we establish notation and survey relevant facts related to the theory of distributions. Then, we introduce the Wigner distribution and state some of its properties. Finally, we introduce the Lie-Trotter product formula and the Mehler-Heine theorem. 
  \begin{figure*}
  \includegraphics[width=\textwidth,height=4cm]{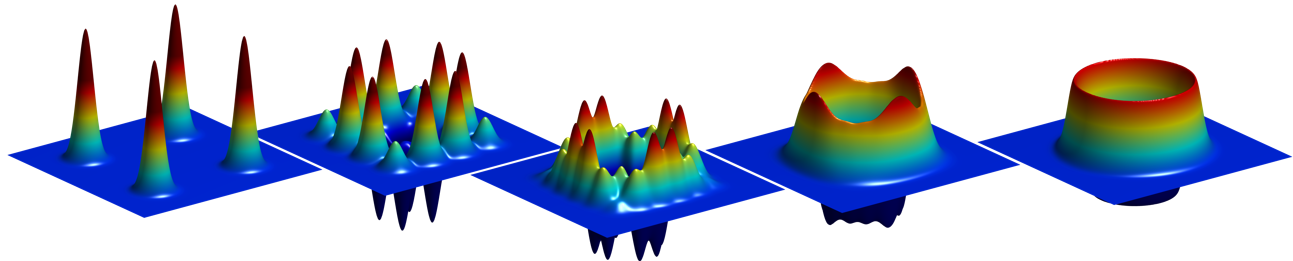}
  \caption{Gaussian-regularized $(\varepsilon = 0.01)$ bivariate quasi-probability distributions for the spin$-1/2$ observables along the $x$ and $y$ directions and the maximally mixed state $\rho = \hat{I}/2$. From left to right: $p_{\text{MH}_1}$, $p_{\text{MH}_3}$, $p_{\text{MH}_5}$, $p_{\text{MH}_{10}}$, and $p_{\text{W}}$.} Each $p_{\text{MH}_m}$ consists of $(m+1)^2$ Dirac delta distributions supported over the grid of points $(1/m)\sum_{i=1}^m\Lambda$.\label{fig:3}
\end{figure*}
 Throughout, we fix  a tuple
($\hat{A}_1,\hdots,\hat{A}_n$) of self-adjoint operators on a finite-dimensional Hilbert space of dimension $d$ and define for $\xi = (\xi_1,\hdots,\xi_n)\in\mathbb{R}^n$ the linear combination
\begin{align*}
    \xi\cdot \hat{A} := \sum_{k= 1}^n\xi_k\hat{A}_k.
\end{align*}
Lastly, we fix a quantum state to be given by a density operator $\rho$, i.e.,\ $\rho\in\mathbb C^{d\times d}$, with $\rho=\rho^\dag \geq 0$ and $\text{tr}(\rho) = 1$.
\subsection{Distributions}
Let $C_0^{\infty}(\mathbb{R}^n)\subseteq\mathcal{S}(\mathbb{R}^n)\subseteq C^{\infty}(\mathbb{R}^n)$ denote the spaces of compactly supported smooth functions, Schwarz functions, and smooth functions on $\mathbb{R}^n$, respectively, and $\mathcal{D}^{'}(\mathbb{R}^n)\supseteq\mathcal{S}^{'}(\mathbb{R}^n)\supseteq\mathcal{E}^{'}(\mathbb{R}^n)$ the corresponding dual spaces of distributions, tempered distributions, and compactly supported distributions on $\mathbb{R}^n$, respectively. The support and singular support of $p\in\mathcal{D}^{'}(\mathbb{R}^n)$ are denoted by supp($p$) and singsupp($p$), respectively.
The $n$-dimensional Fourier transform
\begin{align*}
\mathcal{F}(f)(\xi) = \int_{\mathbb{R}^n}f(x)e^{-ix\cdot\xi}\text{d}x,~~\xi\in\mathbb{R}^n,
\end{align*}
 is an automorphism on $\mathcal{S}(\mathbb{R}^n)$ and
it induces naturally an automorphism on the dual $\mathcal{S}^{'}(\mathbb{R}^n)$. The inverse map is given by Fourier's inversion formula
\begin{align*}
f(x) = \cfrac{1}{(2\pi)^n}\int_{\mathbb{R}^n}\mathcal{F}(f)(\xi) e^{ix\cdot\xi}\text{d}\xi,~~x\in\mathbb{R}^n.
\end{align*}
Next, we state one direction of the Paley-Wiener-Schwartz theorem \cite[Theorem 7.3.1]{hormander1983analysis}, which relates the support properties of a function to analyticity properties of its  Fourier transform. To this end, recall
that the {\em supporting function} of a convex compact set $K\subseteq\mathbb{R}^n$ is 
\begin{align*}
H_K(x) = \sup_{y\in K}\langle x,y\rangle,~~x\in\mathbb{R}^n.
\end{align*}
\textbf{Theorem (Paley-Wiener-Schwartz):} Let $K$ be a convex compact set in $\mathbb{R}^n$. If $f$ is analytic everywhere in $\mathbb{C}^n$ and satisfies
\begin{align*}
\left|f(z)\right|\leq Ce^{H_K(\Im(z))},~~z\in\mathbb{C}^n,
\end{align*}
where $\Im(z)$ denotes the imaginary part of $z$ and $C>0$, then the restriction of $f$ to $\mathbb{R}^n$ is the Fourier transform of a distribution $p\in\mathcal{E}^{'}(\mathbb{R}^n)$ with \begin{align*}
\text{supp}(p)\subseteq K.
\end{align*}

This theorem will be used in Sec. \ref{Sec:III} to study support properties of the distributions $p_{\text{MH}_m}$. A converse statement of the theorem also holds, see \cite[Theorem 7.3.1]{hormander1983analysis}, but will not be needed herein. 

Finally, we say that a sequence of distributions $p_n$ converges to $p$ in $\mathcal{D}^{'}(\mathbb{R}^n)$ as $n\rightarrow\infty$, or simply,  $p_n\rightarrow p\in\mathcal{D}^{'}(\mathbb{R}^n)$ as $n\to\infty$, if for all $\phi\in C_0^{\infty}(\mathbb{R}^n)$, 
\begin{align*}
\lim_{n\rightarrow \infty}\langle p_n,~\phi\rangle =\langle p,~\phi\rangle.
\end{align*}
The same definition applies for $p_n,p\in \mathcal{S}^{'}(\mathbb{R}^n)$ or $\mathcal{E}^{'}(\mathbb{R}^n)$ with respect to test functions $\phi$ taken in $\mathcal{S}(\mathbb{R}^n)$ or $C^{\infty}(\mathbb{R}^n)$, respectively.
 \begin{figure*}
\includegraphics[width=\textwidth,height=4cm]{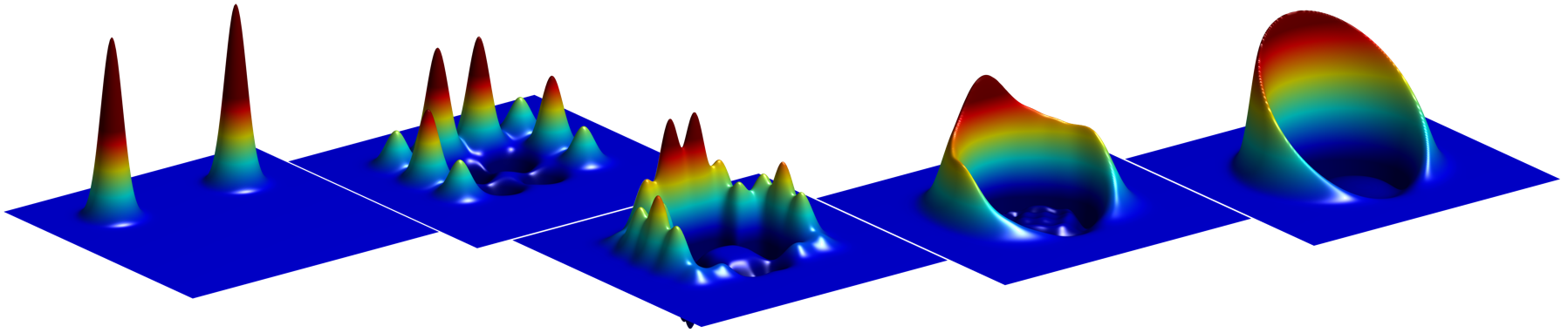}
  \caption{Gaussian-regularized $(\varepsilon = 0.01)$ bivariate quasi-probability distributions for the spin$-1/2$ observables along the $x$ and $y$ directions for the $-\hbar/2$ eigenstate in the $x$ direction. From left to right: $p_{\text{MH}_1}$, $p_{\text{MH}_3}$, $p_{\text{MH}_5}$, $p_{\text{MH}_{10}}$, and $p_{\text{W}}$.} Each $p_{\text{MH}_m}$ consists of $(m+1)^2$ Dirac delta distributions supported over the grid of points $(1/m)\sum_{i=1}^m\Lambda$.\label{fig:4}
\end{figure*}
\subsection{Wigner quasi-probability distribution $p_{\text{W}}$}
The Wigner quasi-probability distribution $p_{\text{W}}$ associated with the observables $\hat{A}_1,\hdots,\hat{A}_n$ and the quantum state $\rho$ is a real-valued distribution in $\mathcal{S}^{'}(\mathbb{R}^n)$  given by 
\begin{align*}
p_{\text{W}} = \cfrac{1}{(2\pi)^n}\mathcal{F}\left(f_{\text{W}}\right),
\end{align*}
where $f_{\text{W}}$ is the quasi-characteristic function 
\begin{align*}
f_{\text{W}}(\xi) = \text{tr}\left(\rho e^{i\xi\cdot {\hat{A}}}\right),~~\xi\in\mathbb{R}^n.
\end{align*}
It was shown in Ref. \cite{10.1063/1.5140632} that $p_{\text{W}}$ is compactly supported, i.e.\ $p_{\text{W}}\in \mathcal{E}^{'}(\mathbb{R}^n),$ and moreover, that
\begin{align*}
\text{supp}(p_{\text{W}})\subseteq R, 
\end{align*}
where the compact convex set $R\subseteq\mathbb{R}^n$ is the \textit{joint numerical range} of the operators $\hat{A}_1,\hdots, \hat{A}_n$. In other words, $R$ is the set of all vectors $a\in\mathbb{R}^n$ with components   $a_i=\text{tr}(\sigma\hat{A}_{i})$ for some density operator $\sigma$. Furthermore, it was shown in Ref. \cite{10.1063/1.5140632} that  
$$
\text{singsupp}(p_{\text{W}})\subseteq S,
$$
where $S$ is the closure of the set of all vectors $a\in\mathbb{R}^n$ with components  $a_i=\text{tr}(\sigma\hat{A}_{i})$ for a subset of density operators $\sigma$, namely, the ones that correspond to non-degenerate eigenstates of $\xi\cdot \hat{A}$. The set $S$ is semi-algebraic (algebraic if $n=2$) and its convex hull is $R$.
\subsection{Lie-Trotter product formula}
 Given any complex-valued matrices $\hat A_1,\hdots,\hat A_n$, then 
 \begin{align}
 \lim_{k\rightarrow \infty}\left(\prod_{i=1}^ne^{\hat A_i/k}\right)^k = e^{\sum_i \hat A_i}.\label{eq:Lie-Trotter}
 \end{align}
 The proof for the case of more than two matrices, i.e., $n>2$, follows verbatim the proof given in
 \cite[Theorem 2.10]{hall2003lie} for two matrices. The formula is also implied by the proof of Lemma $1$ in Appendix \ref{asec:2}. 
 
 The convergence properties of the Lie-Trotter product formula are key to proving the convergence of the Margenau-Hill quasiprobability distributions $p_{\text{MH}_m}$ to the Wigner distribution $p_{\text{W}}$ as $m\rightarrow \infty$. These distributions will be defined and studied in detail in Sec.~\ref{Sec:III}.

\subsection{Mehler-Heine theorem}

The Mehler-Heine theorem describes the asymptotic behavior of the Jacobi polynomials as their degree tends to infinity. These polynomials arise by trotterizing exponentials of spin-$1/2$ operators along orthogonal directions, see Lemma $2$ in Sec.~\ref{sec:4}.
They will be used to elucidate the nature of the convergence of the quasi-probability distributions $p_{\text{MH}_m}$ as $m\rightarrow\infty$.
In particular, we show that the convergence of the quasi-characteristic functions $f_{\text{MH}_m}$ to $f_{\text{W}}$ is a special case of the Mehler-Heine theorem, see Sec.~\ref{sec:4}. 

We first define the Jacobi polynomials
\begin{align}
P_n^{(\alpha,\beta)}(z) = \cfrac{1}{n!}\sum_{k=0}^n\binom{n}{k}c_{n,k}^{\alpha,\beta}\left(\cfrac{z-1}{2}\right)^k,~z\in\mathbb{C},\label{eq:Mehler}
\end{align}
for $\alpha,\beta\in\mathbb{R}$ and $
c_{n,k}^{\alpha,\beta} = (n+\alpha+\beta+1)^{(k)}(\alpha+k+1)^{(n-k)}$.
Then, the Mehler-Heine theorem \cite{szego1975orthogonal} states that 
\begin{align*}
    \lim_{n\rightarrow \infty}n^{-\alpha}P_n^{(\alpha,\beta)}\left(\cos\left(\cfrac{z}{n}\right)\right) =\left(\cfrac{z}{2}\right)^{-\alpha}J_{\alpha}(z),
\end{align*}
 uniformly on compact subsets of $\mathbb{C}$, where $J_{\alpha}(z)$ is the Bessel function of the first kind of order $\alpha$.

  \begin{figure*}
\includegraphics[width=\textwidth,height=4cm]{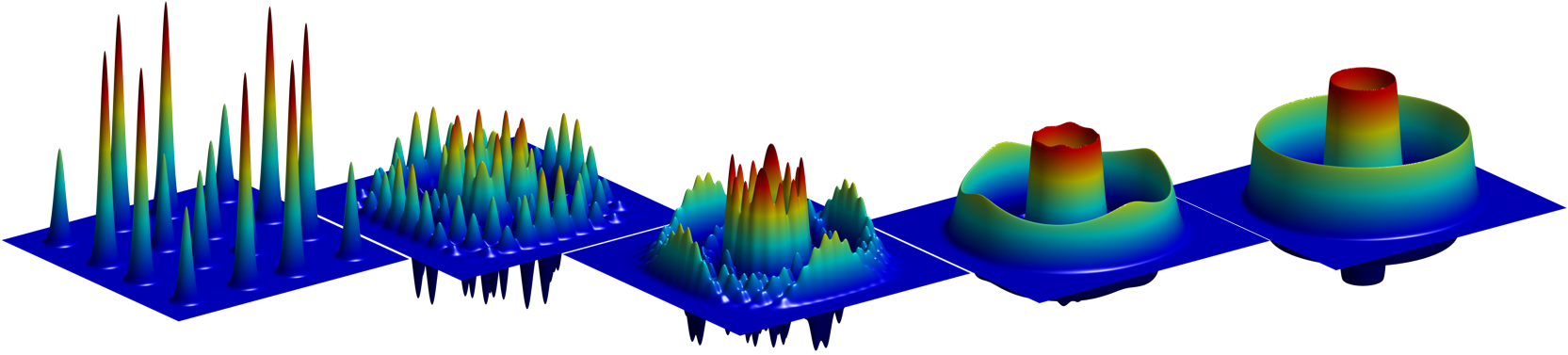}
  \caption{Gaussian-regularized $(\varepsilon = 0.01)$ bivariate quasi-probability distributions for the spin$-3/2$ observables along the $x$ and $y$ directions and the maximally mixed state $\rho = \hat{I}/4$. From left to right: $p_{\text{MH}_1}$, $p_{\text{MH}_3}$, $p_{\text{MH}_5}$, $p_{\text{MH}_{15}}$, and $p_{\text{W}}$.} Each $p_{\text{MH}_m}$ consists of $(3m+1)^2$ Dirac delta distributions supported over the grid of points $(1/m)\sum_{i=1}^m\Lambda$.\label{fig:7}
\end{figure*}

\section{Results}\label{Sec:III}
In this section, we define the Margenau-Hill quasi-probability distributions $p_{\text{MH}_m}$, $m\in\mathbb{N}$, and study their properties. These are discrete signed probability measures associated with the observables $\hat{A}_1,\hdots,\hat{A}_n$ and the quantum state $\rho$. Their marginals along the $i^{\text{th}}$ coordinate coincide with the one induced by the spectral measure of $\hat{A}_i$ and $\rho$, and therefore give the correct probability Law on the measurement outcomes of the observable $\hat{A}_i$. Our main result is that $p_{\text{MH}_m}\rightarrow p_{\text{W}}$ in $\mathcal{E}^{'}(\mathbb{R}^n)$ as $m\rightarrow\infty$, and that the convergence can be upgraded to uniform convergence when the distributions are smeared with an appropriate Schwarz function.

\subsection{Margenau-Hill quasi-probability distributions $p_{\text{MH}_m}$}\label{sec:MH}
We define the Margenau-Hill quasi-probability distribution $p_{\text{MH}_m}$ of order $m\in\mathbb{N}$ associated with the observables $\hat{A}_1,\hdots,\hat{A}_n$ and the quantum state $\rho$ to be 
\begin{align*}
p_{\text{MH}_{m}} = \cfrac{1}{(2\pi)^n}\mathcal{F}\left(f_{\text{MH}_m}\right),
\end{align*}
where $f_{\text{MH}_m}$ is the quasi-characteristic function 
\begin{align}
    f_{\text{MH}_m}(\xi) = \cfrac{1}{n!}~\text{tr}\left(\rho\sum_{\pi\in S_n}\left(\prod_{k=1}^ne^{i\frac{\xi_{\pi(k)}}{m}\hat{A}_{\pi(k)}}\right)^m\right),\label{eq:MH}
\end{align}
and $S_n$ is the symmetric group. That is, $S_n$ is the set of all permutations of $n$ elements. The products inside the summation are ordered from left to right, starting from the lowest index $k=1$. That is, 
\begin{align*}
\prod_{k=1}^ne^{i\frac{\xi_{\pi(k)}}{m}\hat{A}_{\pi(k)}} = e^{i\frac{\xi_{\pi(1)}}{m}\hat{A}_{\pi(1)}}e^{i\frac{\xi_{\pi(2)}}{m}\hat{A}_{\pi(2)}}\hdots e^{i\frac{\xi_{\pi(n)}}{m}\hat{A}_{\pi(n)}}.
\end{align*}The reason why the definition of $f_{\text{MH}_m}$ includes a summation over all possible permutations $\pi\in S_n$ is to ensure that the distributions $p_{\text{MH}_m}$ are real-valued for all $m\in\mathbb{N}.$ For example, when $n=2$ and $m=1$, we recover the Margenau-Hill quasi-characteristic function
\begin{align*}
f_{\text{MH}_m}(\xi_1,\xi_2) = \text{tr}\left(\rho \cfrac{e^{i\xi_1\hat{A}_1}e^{i\xi_2\hat{A}_2}+e^{i\xi_2\hat{A}_2}e^{i\xi_1\hat{A}_1}}{2}\right),
\end{align*}
studied in \cite{Cohen1986-COHJWD} for the spin$-1/2$ operators $\hat{A}_1 =\hat{S}_1$ and $\hat{A}_2=\hat{S}_2$. In what follows, we state and prove properties of $p_{\text{MH}_m}$, starting with the most basic ones.\\\\
 \textbf{Proposition $1$:} The distributions $p_{\text{MH}_m}$ are tempered and real-valued, i.e., for all $m\in\mathbb{N}$, $p_{\text{MH}_m}\in \mathcal{S}^{'}(\mathbb{R}^n)$ and
 \begin{align*}
\langle\overline{p_{\text{MH}_m}},\phi \rangle = \langle p_{\text{MH}_m},\phi\rangle,~~\forall\phi\in\mathcal{S}(\mathbb{R}^n).
 \end{align*}
\textbf{Proof:}
The functions $f_{\text{MH}_m}$ are continuous in $\xi\in\mathbb{R}^n$, and uniformly bounded since
\begin{align*}
    |f_{\text{MH}_m}(\xi)|&\leq \cfrac{1}{n!}~\sum_{\pi\in S_n}\text{tr}\left(\left|\rho\left(\prod_{k=1}^ne^{i\frac{\xi_{\pi(k)}}{m}\hat{A}_{\pi(k)}}\right)^m\right|\right),\\
    &= \cfrac{1}{n!}~\sum_{\pi\in S_n}\text{tr}\left(\left|\rho\right|\right) = \text{tr}(\rho) = 1,~\forall m\in\mathbb{N},
\end{align*}
where $|A| := \sqrt{AA^{\dag}}$. Thus, $\forall m\in\mathbb{N}$, the maps
\begin{align*}
    u_m:~\phi(\xi)\mapsto \int_{\mathbb{R}^n}f_{\text{MH}_m}(\xi)\phi(\xi)\text{d}\xi,\;\;\phi(\xi)\in\mathcal{S}(\mathbb{R}^n),
\end{align*}
are continuous linear forms on $\mathcal{S}(\mathbb{R}^n)$. That is, the distributions $u_m$ are tempered. Consequently, their images under the Fourier transform, and therefore
\begin{align*}
    p_{\text{MH}_m} =\cfrac{1}{(2\pi)^n}\mathcal{F}(u_m),
\end{align*}
are also tempered. Finally, since
$$
\overline{f_{\text{MH}_m}(\xi)}=f_{\text{MH}_m}(-\xi),~~\forall m\in\mathbb{N},
$$ 
we have 
\begin{align*}
&\langle\overline{p_{\text{MH}_m}},\phi \rangle :=\overline{\langle p_{\text{MH}_m},\overline{\phi}\rangle}  = \cfrac{1}{(2\pi)^n}\overline{\langle u_m,~\mathcal{F}\left(\overline{\phi}\right) \rangle} \\
   &= \cfrac{1}{(2\pi)^n}\int_{\mathbb{R}^n}f_{\text{MH}_m}(-\xi)\mathcal F({\phi})(-\xi)\text{d}\xi =  \langle p_{\text{MH}_m},\phi\rangle,
\end{align*}
which completes the proof.$\hfill\blacksquare$\\

As done for the Wigner distribution $p_{\text{W}}$ in Ref. \cite{10.1063/1.5140632}, we demonstrate next how the Paley-Wiener-Schwarz theorem can be used to prove that the Margenau-Hill distributions $p_{\text{MH}_m}$ are compactly supported for all $m\in\mathbb{N}$. To this end, define the set 
$$
\Lambda := \sigma(\hat{A}_1)\times\sigma(\hat{A}_2)\times\hdots\times\sigma(\hat{A}_n),
$$
where $\sigma(\hat{A}_k)$ is the spectrum of $\hat{A}_k$, $k=1,\hdots,n$. The set $\Lambda$ consists of all tuples of eigenvalues and is the classical support that is expected from a joint probability distribution for the simultaneous measurement outcomes of the observables $\hat{A}_1,\hdots,\hat{A}_n$. The convex hull of $\Lambda$, denoted by $\text{conv}(\Lambda)$, is the \textit{free numerical range} of the observables $\hat{A}_1,\hdots,\hat{A}_n$, which is the set of all vectors in $\mathbb{R}^n$ with the $i^{\text{th}}$ component being equal to $\text{tr}(\rho_i\hat{A}_i)$ for some density matrix $\rho_i$. This set contains the \textit{joint numerical range} $R$, which in turn contains the support of the Wigner distribution $p_{\text{W}}$.\\\\
\begin{figure*}
\includegraphics[width=\textwidth,height=4cm]{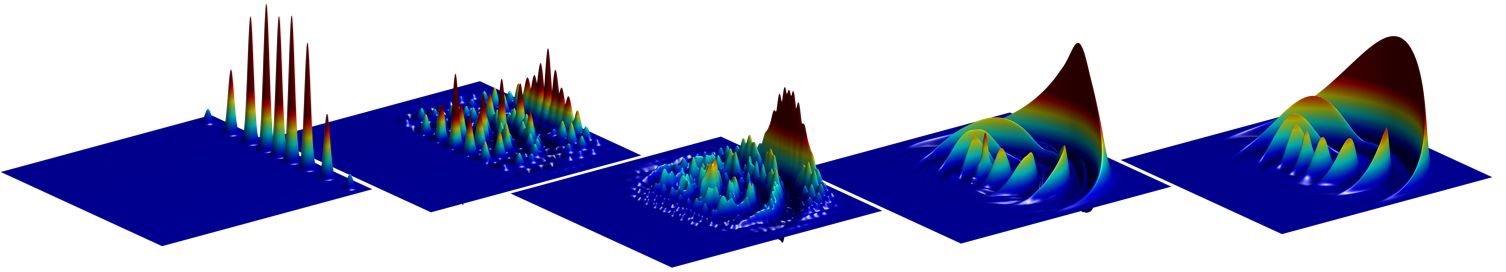}
  \caption{Gaussian-regularized $(\varepsilon = 0.1)$ bivariate quasi-probability distributions for the spin$-4$ observables along the $x$ and $y$ directions for the $+4\hbar$ eigenstate  along the $y$ direction. From left to right: $p_{\text{MH}_1}$, $p_{\text{MH}_2}$, $p_{\text{MH}_3}$, $p_{\text{MH}_{10}}$, and $p_{\text{W}}$.} Each $p_{\text{MH}_m}$ consists of $(8m+1)^2$ Dirac delta distributions supported over the grid of points $(1/m)\sum_{i=1}^m\Lambda$.\label{fig:5}
\end{figure*}\noindent
\textbf{Proposition $2$:}  For all $m\in\mathbb{N}$, the distributions $p_{\text{MH}_m}$ are compactly supported, i.e., $p_{\text{MH}_m}\in\mathcal{E}^{'}(\mathbb{R}^n)$, and moreover,
$$
\text{supp}(p_{\text{MH}_m})\subseteq \text{conv}(\Lambda),
$$
where $\text{conv}(\Lambda)$ is the convex hull of $\Lambda$.\\\\
\textbf{Proof:} 
    The set $K:=\text{conv}(\Lambda)$ is a convex solid in $\mathbb{R}^n$ with vertex set 
 $$
    \mathcal{V}(K) = \{v\in\mathbb{R}^n~|~v_k \in\{ \lambda_{\text{min}}(\hat{A}_k),\; \lambda_{\text{max}}(\hat{A}_k)\}, \;\forall k\},
    $$
    where $\lambda_{\rm min}$ and $\lambda_{\rm max}$ denote the minimal and maximal eigenvalues.
   Thus, the supporting function of $K$ is
    \begin{align*}
        H_{K}(x) = \max_{v\in\mathcal{V}(K)}\langle v,x\rangle. 
    \end{align*}
   By Proposition $1$,  $p_{\text{MH}_m}\in\mathcal{S}^{'}(\mathbb{R}^n)$ for all $m\in\mathbb{N}$. Then by Fourier's inversion formula, we have 
    \begin{align*}
       \mathcal{F}(p_{\text{MH}_m}) = \cfrac{1}{(2\pi)^n}\mathcal{F}(\mathcal{F}(f_{\text{MH}_m})) = g_{\text{MH}_m},
    \end{align*}
 where $g_{\text{MH}_m}(\xi) = f_{\text{MH}_m}(-\xi)$.
Finally, observe that the function $g_{\text{MH}_m}(z)$, $z\in\mathbb{C}^n$, is analytic everywhere in $\mathbb{C}^n$ and satisfies the estimate
    \begin{align*}
        |g_{\text{MH}_m}(z)| &\leq \cfrac{1}{n!}~\sum_{\pi\in S_n}\text{tr}\left(\left|\rho\left(\prod_{k=1}^ne^{i\frac{-z_{\pi(k)}}{m}\hat{A}_{\pi(k)}}\right)^m\right|\right),\\
        & \leq \cfrac{e^{\max_{v\in \mathcal{V}(K)}\langle v,\Im(z)\rangle}}{n!}\sum_{\pi\in S_n}\text{tr}\left(\rho\right) =e^{H_{K}(\Im(z))}, 
    \end{align*}
    for all $m\in\mathbb{N}$. Detailed steps for the last inequality are given in Appendix \ref{asec:1}. By the Paley-Wiener-Schwarz theorem, this completes the proof. $\hfill\blacksquare$\\
    
    The above is a powerful approach that can be extended straightforwardly to bounded self-adjoint operators on an infinite-dimensional Hilbert space. Therein, $\Lambda$ need not be a finite set.  In the current finite-dimensional setting, however, $\Lambda$ is always a finite set. Because of this, Proposition $2$ can be significantly refined, and the support can be characterized more accurately as follows.\\\\
\textbf{Proposition $3$:} For all $m\in\mathbb{N}$, $p_{\text{MH}_m}$ is a finite affine combination of  Dirac delta distributions with\begin{align*}
    \text{singsupp}(p_{\text{MH}_m}) = \text{supp}(p_{\text{MH}_m}) \subseteq \cfrac{1}{m}\sum_{i = 1}^m\Lambda,
\end{align*}
where the summation is in the sense of Minkowski.\\\\
\textbf{Proof:}
  Let $f^{\ast m}:=f\ast f\ast\hdots\ast f$ denote the convolution of $f$ with itself $m$ times. Then, explicit formal computation shows that for all $m\in\mathbb{N}$, $x\in\mathbb{R}^n$,
    \begin{figure*}
\includegraphics[width=\textwidth,height=4cm]{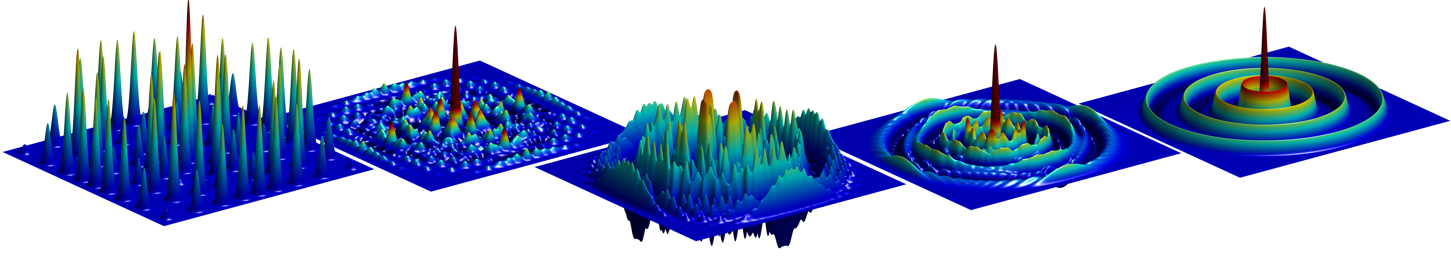}
  \caption{Gaussian-regularized $(\varepsilon = 0.1)$ bivariate quasi-probability distributions for the spin$-4$ observables along the $x$ and $y$ directions and the maximally mixed state $\rho = \hat{I}/9$. From left to right: $p_{\text{MH}_1}$, $p_{\text{MH}_2}$, $p_{\text{MH}_3}$, $p_{\text{MH}_{5}}$, and $p_{\text{W}}$.} Each $p_{\text{MH}_m}$ consists of $(8m+1)^2$ Dirac delta distributions supported over the grid of points $(1/m)\sum_{i=1}^m\Lambda$.\label{fig:6}
\end{figure*}
\begin{align*}
    &p_{\text{MH}_m}(x) = \cfrac{1}{(2\pi)^n}\int_{\mathbb{R}^n}f_{\text{MH}_m}(\xi)e^{-i x\cdot\xi}\text{d}\xi\\
    & = \cfrac{1}{n!(2\pi)^n}\sum_{\pi\in S_n}~\text{tr}\left(\rho\int_{\mathbb{R}^n}\left(\prod_{k=1}^ne^{i\frac{\xi_{\pi(k)}}{m}\hat{A}_{\pi(k)}}\right)^m\hspace{-3mm}e^{-i x\cdot\xi}\text{d}\xi\right),
\end{align*}
where 
\begin{align*}
    &\int_{\mathbb{R}^n}\left(\prod_{k=1}^ne^{i\frac{\xi_{\pi(k)}}{m}\hat{A}_{\pi(k)}}\right)^me^{-i x\cdot\xi}\text{d}\xi\\&
    = \left(\int_{\mathbb{R}^n}\left(\prod_{k=1}^ne^{i\frac{\xi_{\pi(k)}}{m}\hat{A}_{\pi(k)}}\right)e^{-i x\cdot\xi}\text{d}\xi \right)^{\ast m}\\
    &=\left(\prod_{k=1}^n\left(\int_{\mathbb{R}}e^{i\frac{\xi_{\pi(k)}}{m}\hat{A}_{\pi(k)}}e^{-i x_{\pi(k)}\xi_{\pi(k)}}\text{d}\xi_{\pi(k)}\right) \right)^{\ast m}\\
    &=\left(\prod_{k=1}^n\hat{U}_{\pi(k)}\hspace{-1mm}\left(\int_{\mathbb{R}}\hspace{-1mm}e^{-i\xi_{\pi(k)}(x_{\pi(k)}-\frac{\hat D_{\pi(k)}}{m} )}\text{d}\xi_{\pi(k)}\right)\hspace{-0.8mm}\hat{U}^{\dag}_{\pi(k)} \right)^{\ast m}\\&= \left(\prod_{k=1}^n\hat{U}_{\pi(k)}\hat{E}_{\pi(k)}\hat{U}^{\dag}_{\pi(k)} \right)^{\ast m},
\end{align*}
and  $\hat{U}_{\pi(k)}\hat{D}_{\pi(k)}\hat{U}^{\dag}_{\pi(k)}$ is the eigendecomposition of $\hat{A}_{\pi(k)} $. The diagonal matrix $\hat{E}_{\pi(k)}$, given by 

$$
\hat{E}_{\pi(k)} =\int_{\mathbb{R}}e^{-i\xi_{\pi(k)}\left(x_{\pi(k)}-\frac{\hat D_{\pi(k)}}{m} \right)}\text{d}\xi_{\pi(k)}, 
$$
has in its $i^{\text{th}}$ diagonal entry the Dirac delta distribution
$$
2\pi\cdot\delta\left(x_{\pi(k)}-\cfrac{[\hat{D}_{\pi(k)}]_{ii}}{m}\right),
$$ where $[\hat{D}_{\pi(k)}]_{ii}$ is the $i^{\text{th}}$ diagonal entry of $\hat{D}_{\pi(k)}$. This implies that 
\begin{align*}
&\text{singsupp}\left(\prod_{k=1}^n\hat{U}_{\pi(k)}\hat{E}_{\pi(k)}\hat{U}^{\dag}_{\pi(k)} \right)\\&=\text{supp}\left(\prod_{k=1}^n\hat{U}_{\pi(k)}\hat{E}_{\pi(k)}\hat{U}^{\dag}_{\pi(k)} \right) \subseteq \Lambda/m.
\end{align*}
By performing the convolution $m$ times, the resulting support for $p_{\text{MH}_m}$  is contained in the set $\Lambda/m$ added to itself $m$ times. Thus, the distribution $p_{\text{MH}_m}$ is a linear combination of Dirac delta distributions supported in $(1/m)\sum_{i=1}^m\Lambda$. To show that this linear combination is affine, i.e., that the coefficients of the combination add up to $1$, it is enough to note that 
 \begin{align*}
     \langle p_{\text{MH}_m},1\rangle &=\cfrac{1}{(2\pi)^n}\int_{\mathbb{R}^n}f_{\text{MH}_m}(\xi)\mathcal{F}(1)~\text{d}\xi
     \\& = f_{\text{MH}_m}(0) = 1,
 \end{align*}
 which completes the proof. \hfill $\blacksquare$\\

Thus, the Margenau-Hill distribution of any order is a discrete signed probability measure associated with the observables $\hat{A}_1,\hdots,\hat{A}_n$ and the quantum state $\rho$. Finally, we verify that the marginals of $p_{\text{MH}_m}$ along any  principal axis gives the correct probability law on the measurement outcomes on the corresponding observable.\\\\
\textbf{Proposition $4$:} The $j^{\text{th}}$ marginal of $p_{\text{MH}_m}$ is given by
$$
\int_{\mathbb{R}^{n-1}}p_{\text{MH}_m}\text{d}x_{\backslash j}= \sum_{s = 1}^d\langle\psi_j(s)|\rho|\psi_j(s)\rangle  \delta(x_j-\lambda_j(s)), 
$$
where $\text{d}x_{\backslash j}:=\text{d}x_1\hdots\text{d}x_{j-1}\text{d}x_{j+1}\hdots\text{d}x_n$ and $\lambda_j(s)$, $|\psi_j(s)\rangle$ are the corresponding eigenvalues and eigenvectors of $\hat{A}_j$, respectively, with $s = 1,\hdots,d$.  \\\\
\textbf{Proof:} Starting from the left-hand side of the above,
\begin{align*}
    &\cfrac{1}{(2\pi)^n}\left.\int_{\mathbb{R}^{n-1}}\mathcal{F}(f_{\text{MH}_m})e^{i\sum_{k\neq j}^nx_k\xi_k}\text{d}x_{\backslash j}\right\vert_{\xi_k=0,\,\forall k\neq j}\\
    &= \left.\cfrac{1}{2\pi}\int_{\mathbb{R}}f_{\text{MH}_m}(\xi)e^{-ix_j\xi_j}\text{d}\xi_j\right\vert_{\xi_k=0,\,\forall k\neq j}\\
    &= \cfrac{1}{2\pi}\int_{\mathbb{R}}\text{tr}\left(\rho e^{i\xi_j\hat{A}_j}\right)e^{-ix_j\xi_j}\text{d}\xi_j\\
    &= \text{tr}\left(\rho \hat{U}_{j}\left(\cfrac{1}{2\pi}\int_{\mathbb{R}}e^{-i\xi_j(x_j-\hat{D}_j)}\text{d}\xi_j\right)\hat{U}^{\dag}_{j}\right)\\
    &= \sum_{s = 1}^d\langle\psi_j(s)|\rho|\psi_j(s)\rangle  \delta(x_j-\lambda_j(s)),
\end{align*}
where $\hat{A}_j = \hat{U}_j\hat{D}_j\hat{U}_j^{\dag}$ is the eigendecomposition of $\hat{A}_j$, and $\lambda_j(s)$, $|\psi_j(s)\rangle$ as claimed. \hfill $\blacksquare$

\subsection{Convergence to the Wigner distribution $p_{\text{W}}$}
Herein, we prove that $p_{\text{MH}_m}\rightarrow p_{\text{W}}$ in $\mathcal{E}^{'}(\mathbb{R}^n)$ as $m\rightarrow\infty$, and that the convergence can be upgraded to uniform convergence when the distributions are smeared with an appropriate Schwarz function. \\\\
\textbf{Proposition  $5$:} The Margenau-Hill quasi-probability distributions $p_{\text{MH}_m}$ converge to the Wigner distribution $p_{\text{W}}$ in $\mathcal{E}^{'}(\mathbb{R}^n)$ as $m\rightarrow\infty$.\\\\
\textbf{Proof:} By definition, we must show that 
     \begin{align*}
\lim_{m\rightarrow\infty}\langle p_{\text{MH}_m},~\phi\rangle =\langle p_{\text{W}},~\phi\rangle,~~\forall \phi\in C^{\infty}(\mathbb{R}^n).
        \end{align*}
To that end, let $\chi\in C_0^{\infty}(\mathbb{R}^n)$ be a cutoff function equal to $1$ on a neighborhood of $K =\text{conv}(\Lambda)$ and let 
$$
\psi = \chi\phi\in C_{0}^{\infty}(\mathbb{R}^n),
$$ for any $\phi\in C^{\infty}(\mathbb{R}^n)$, then
        \begin{align} 
             \langle p_{\text{MH}_m},~\phi\rangle &= \langle p_{\text{MH}_m},~\psi\rangle\nonumber\\
             &= \cfrac{1}{(2\pi)^n}\langle \mathcal{F}(f_{\text{MH}_m}),~\psi\rangle\nonumber \\
             &= \cfrac{1}{(2\pi)^n}\int_{\mathbb{R}^n}f_{\text{MH}_m}(\xi)\mathcal{F}(\psi)(\xi)\text{d}\xi.\label{integrand}
        \end{align}
By the Lie-Trotter product formula \eqref{eq:Lie-Trotter}, we have the pointwise convergence 
    \begin{align*}
        \lim_{m\rightarrow \infty}f_{\text{MH}_m}(\xi) = \text{tr}\left(\rho e^{i\xi\cdot A}\right)=f_{\text{W}}(\xi).
    \end{align*}
     Thus, the integrand in Eq. \eqref{integrand} converges pointwise to $f_{\text{W}}(\xi)\mathcal{F}(\psi)(\xi)$ and is bounded in absolute value by $|\mathcal{F}(\psi)(\xi)|\in \mathcal{S}(\mathbb{R}^n)$ for all $m\in\mathbb{N}$. By the dominated convergence theorem we get 
        \begin{align*}
         &\lim_{m\rightarrow \infty} \langle p_{\text{MH}_m},~\phi\rangle  = \cfrac{1}{(2\pi)^n}\int_{\mathbb{R}^n}f_{\text{W}}(\xi)\mathcal{F}(\psi)(\xi)\text{d}\xi\\
         &= \cfrac{1}{(2\pi)^n}\langle \mathcal{F}(f_{\text{W}}),~\psi\rangle = \langle p_{\text{W}},~\psi\rangle=\langle p_{\text{W}},~\phi\rangle ,
        \end{align*}
        where the last equality follows from the fact that $\text{supp}(p_{\text{W}})\subseteq R \subseteq K$. \hfill $\blacksquare$\\\\
        Thus, we have shown that, while the general Wigner distribution $p_{\text{W}}$ need not be a quasi-probability measure for the associated observables, it is not far from being one. More precisely, Proposition $5$ implies that for any $\epsilon >0$ and $\phi\in C^{\infty}(\mathbb{R}^n)$, there will exist a discrete signed quasi-probability measure $\mu$ such that 
$$
\left|\langle p_{\text{W}},~\phi\rangle - \int_{\mathbb{R}^n}\phi\text{d}\mu\right|\leq \epsilon,
$$
with the measure $\mu$ being a Margenau-Hill quasi-probability distribution $p_{\text{MH}_m}$ for sufficiently large $m$.\\

       Finally, the convergence result in Proposition $5$ can be upgraded to uniform convergence if the distributions are smeared, i.e., convolved, with an appropriate Schwarz function. To prove this,  the following lemma is needed.\\\\
       \textbf{Lemma $1$:} The Margenau-Hill quasi-characteristic functions $f_{\text{MH}_m}$ converge to the Wigner quasi-characteristic function $f_{\text{W}}$ as $m\rightarrow \infty$ uniformly on compact subsets of $\mathbb{C}^n$.\\\\
       \textbf{Proof:} The proof is given in Appendix \ref{asec:2}.\hfill$\blacksquare$\\
       \begin{figure}
\includegraphics[scale = 0.34]{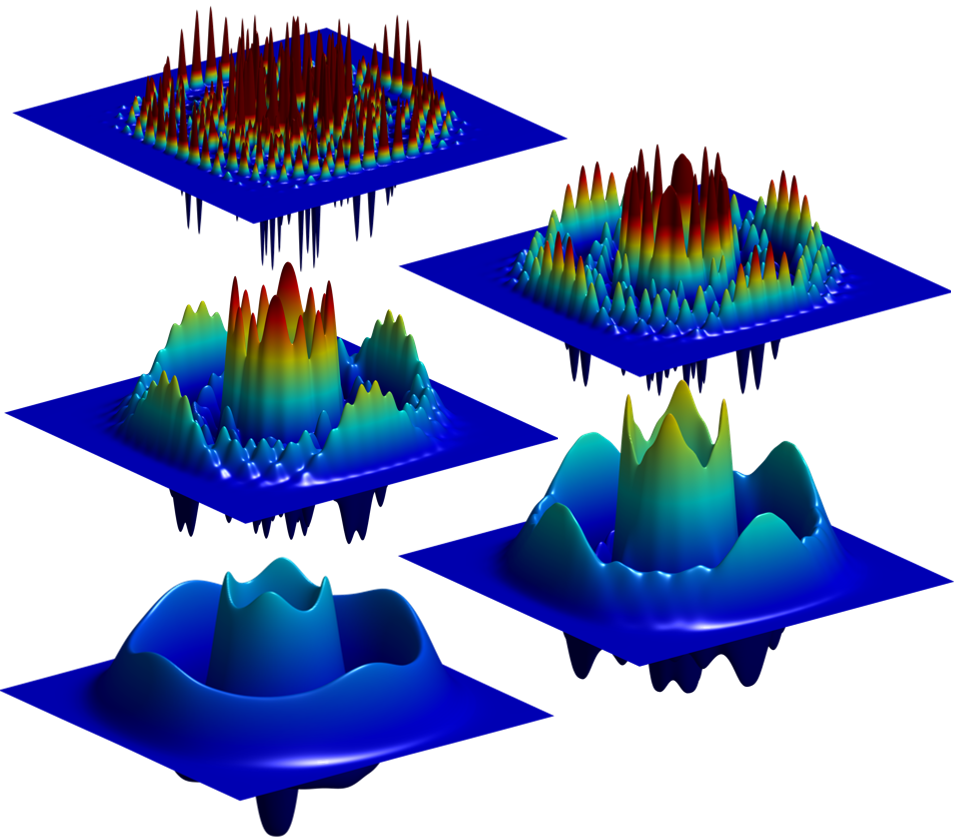}
\caption{Dependence of the resolution of the features of the gaussian-regularized Margenau-Hill distribution $p_{\text{MH}_5}$ on the regularizing parameter $\varepsilon$, for the spin$-3/2$ observables along the $x$ and $y$ directions for the maximally mixed state $\rho= \hat{I}/4$. From top to bottom: $\varepsilon = 0.001$, $0.005$, $0.01$, $0.02$, $0.05$. The distribution $p_{\text{MH}_5}$ consists of $16^2$ Dirac delta distributions supported over the grid of points $1/5\sum_{i=1}^5\lambda$. They are best visualized for $\varepsilon = 0.001$.} 
    \label{fig:res}
\end{figure}

        Evidently, Lemma $1$ refers to the analytic extensions of $f_{\text{MH}_m}$ and $f_{\text{W}}$ to all $\mathbb{C}^n$. In particular, the original functions $f_{\text{MH}_m}$ converge to $f_{\text{W}}$ as $m\rightarrow \infty$ uniformly on compact subsets of $\mathbb{R}^n$.\\\\
       \textbf{Proposition $6$:} If $\psi\in\mathcal{S}(\mathbb{R}^n)$ satisfies $\mathcal{F}(\psi)\in C_{0}^{\infty}(\mathbb{R}^n)$, then $\psi\ast p_{\text{MH}_m} \rightarrow \psi\ast p_{\text{W}}$ uniformly as $m\rightarrow \infty$.\\\\
       \textbf{Proof:}
Recall that since $\psi$ is smooth and the distributions $p_{\text{MH}_m}$ and $p_{\text{W}}$ are tempered, then $\psi\ast p_{\text{MH}_m}$ and $\psi\ast p_{\text{W}}$ are also smooth. Let $\text{supp}(\mathcal{F}(\psi))\subseteq K$ compact, then 
\begin{align}
    &|\psi\ast p_{\text{W}} - \psi\ast p_{\text{MH}_m}|(x)\nonumber\\
    &= |\int_{\mathbb{R}^n}\mathcal{F}(\psi)(\xi)\left(f_{\text{W}}(-\xi)-f_{\text{MH}_m}(-\xi)\right)e^{-i\xi\cdot x}~\text{d}\xi|\nonumber\\
    &\leq \int_{K}|\mathcal{F}(\psi)(\xi)||f_{\text{W}}(-\xi)-f_{\text{MH}_m}(-\xi)|\text{d}\xi \nonumber\\
    &\leq \left(\int_{K}|\mathcal{F}(\psi)(\xi)|\text{d}\xi\right)\cdot\sup_{\xi\in K}|f_{\text{W}}(-\xi)-f_{\text{MH}_m}(-\xi)|\nonumber\\
    &\leq \left(\int_{K}|\mathcal{F}(\psi)(\xi)|\text{d}\xi\right)\cdot\sup_{\xi\in -K}|f_{\text{W}}(\xi)-f_{\text{MH}_m}(\xi)|,\label{uniform}
\end{align}
where the last integral is finite as $\mathcal{F}(\psi)\in\mathcal{S}(\mathbb{R}^n)$ and therefore integrable. By Lemma $1$, $f_{\text{MH}_m}\rightarrow f_{\text{W}}$  as $m\rightarrow \infty$ uniformly on compact subsets of $\mathbb{R}^n$. Thus, 
$$
\sup_{\xi\in -K}|f_{\text{W}}(\xi)-f_{\text{MH}_m}(\xi)|\rightarrow 0,
$$
as $m\rightarrow \infty$. Taking the supremum over all $x\in\mathbb{R}^n$ on both sides of Eq. \eqref{uniform} implies that 
\begin{align*}
    \sup_{x\in\mathbb{R}^n}|\psi\ast p_{\text{W}} - \psi\ast p_{\text{MH}_m}| \rightarrow 0,
\end{align*}
as $m\rightarrow \infty$, which completes the proof.$\hfill\blacksquare$\\\\
We illustrate in Figs.~\ref{fig:3}--\ref{fig:6} the convergence of the Margenau-Hill quasi-probability distributions $p_{\text{MH}_m}$ to the corresponding Wigner distribution $p_{\text{W}}$ for pairs of spin operators along the $x$ and $y$ directions and various quantum states. The figures display smoothed versions of the distributions obtained by convolving with a Gaussian function.
This is effected by multiplying the quasi-characteristic function $f$ by the decaying exponential $e^{-\varepsilon\xi^2}$, where the regularizing parameter $\varepsilon$ is tuned for best visualization, as explained in Ref.~\cite{10.1063/1.5140632}. Specifically, when $\varepsilon$ is too large, the features of the distribution $p$ are wiped out, and when $\varepsilon$ is to small, singularities manifest with exceedingly large values. An example of the effect of the regularizing parameter $\varepsilon$ on the resolution of the plots is shown in Fig. \ref{fig:res}.

     \section{Spin$-1/2$}\label{sec:4}
In this section, we specialize to quasi-probability distributions for a pair of spin$-1/2$ observables by setting
$$
\hat{A}_1 = \hat{S}_1:=\hat{S}\cdot\hat{n}_1,~~~\hat{A}_2 =\hat{S}_2:=\hat{S}\cdot\hat{n}_2,
$$ where $\hat{S}$ is the spin$-1/2$ operator and $\hat{n}_1$ and $\hat{n}_2$ are orthogonal directions in the Bloch sphere. The directions are given by the unit vectors 
\begin{align*}
    \hat{n}_i &= \sin(\theta_i)\cos(\phi_i)\hat{x}+\sin(\theta_i)\sin(\phi_i)\hat{y}+\cos(\theta_i)\hat{z},
\end{align*} where $\hat{n}_1\cdot\hat{n}_2=0$, for some $\theta_i,~\phi_i\in\mathbb{R}$, $i \in \{1,2\}$. First, we recall some properties of $\hat{S}_1$ and $\hat{S}_2$ as well as their Wigner distribution studied in Refs. \cite{Cohen1986-COHJWD,10.1063/1.5140632}. Then we compute for all $m\in\mathbb{N}$ closed-form expressions for the Margenau-Hill quasi-characteristic functions $f_{\text{MH}_m}$ and quasi-probability distributions $p_{\text{MH}_m}$. As a byproduct, we elucidate the nature of the convergence of $f_{\text{MH}_m}$ to $f_{\text{W}}$ as $m\rightarrow\infty$ by relating it to a special case of the Mehler-Heine theorem. For simplicity {\color{black} of exposition}, we replace the spin$-1/2$ values $\pm\hbar/2$ by $\pm 1$.
\begin{figure}[t!]
\includegraphics[scale = 0.5]{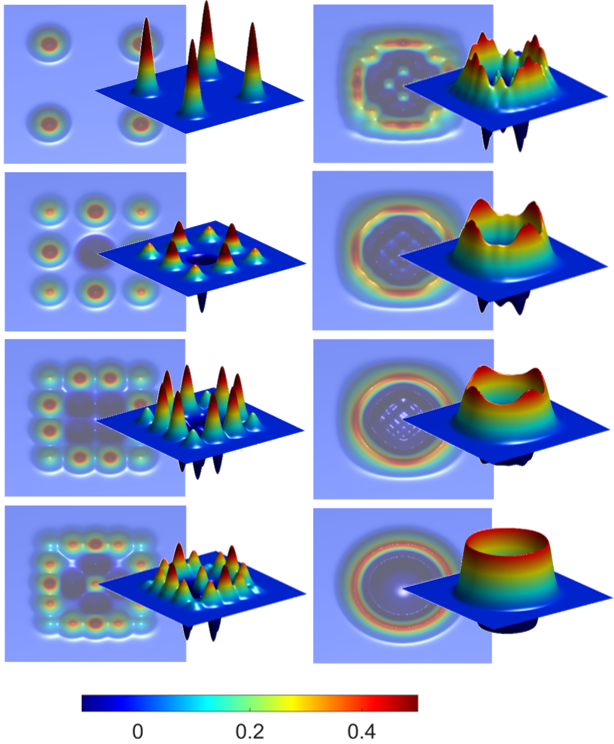}
\caption{Evolution of the support of the Margenau-Hill distribution $p_{\text{MH}_m}$ for the spin$-1/2$ observables along the $x$ and $y$ directions and the maximally mixed state $\rho= \hat{I}/2$. From top to bottom: (left) $p_{\text{MH}_1}$, $p_{\text{MH}_2}$, $p_{\text{MH}_3}$, $p_{\text{MH}_4}$,
(right) $p_{\text{MH}_{6}}$, $p_{\text{MH}_{8}}$, $p_{\text{MH}_{11}}$, $p_{\text{W}}$. The distribution $p_{\text{MH}_m}$ consists of $(m+1)^2$ Dirac delta distributions supported over the grid of points $1/m\sum_{i=1}^m\lambda$.} 
    \label{fig:resspin9}
\end{figure}
\subsection{Properties of $\hat{S}_1$ and $\hat{S}_2$}
The operators $\hat{S}_1$ and $\hat{S}_2$, represented in the $\hat{S}_z:=\hat{S}\cdot\hat{z}$ eigenbasis, are given by
\begin{align*}
\hat{S}_j &= \begin{bmatrix}
        \cos(\theta_j)&\phantom{-}e^{-i\phi_j}\sin(\theta_j)\\
        e^{i\phi_j}\sin(\theta_j)&-\cos(\theta_j)
    \end{bmatrix},~j \in \{1,2\},
\end{align*}
and satisfy various properties summarized below.  \\\\
\textbf{Proposition $7$:} Let $\hat{I}$ denote the identity matrix. Then, for all $\xi_1,~\xi_2\in\mathbb{R}$, $j,k\in\{1,2\}$, the following properties hold:
\begin{align*}
    \hat{S}_j^2 &= \hat{I},~~\text{det}(\hat{S}_j)=-1,~~\text{tr}(\hat{S}_j) = 0,\\
     \left[\hat{S}_j,\hat{S}_k\right] &= 2i\hat{S}\cdot(\hat{n}_j\times\hat{n}_k),~~\{\hat{S}_j,\hat{S}_k\} = 2\left(\hat{n}_j\cdot\hat{n}_k\right)\hat{I},\\
e^{i(\xi_1\hat{S}_1+\xi_2\hat{S}_2)} &= \cos(\|\xi\|)\hat{I} + i(\xi_1\hat{S}_1+\xi_2\hat{S}_2)\frac{\sin(\|\xi\|)}{\|\xi\|},
\end{align*}
where $\|\xi\| =\sqrt{\xi_1^2+\xi_2^2}$. The symbol $\times$ denotes the cross product, and $\left[\cdot\,,\cdot\right]$ and $\{\cdot\,,\cdot\}$ denote the commutator and anti-commutator brackets, respectively.\\\\
\textbf{Proof:} The proof is given in Appendix \ref{asec:3}.\hfill$\blacksquare$\\\\
Next, recall that the quantum state $\rho$ of a spin$-1/2$ particle can be represented in the $\hat{S}_z$ eigenbasis by 
$$
\rho = \cfrac{1}{2}\begin{bmatrix}
    1+z& x-iy\\
    x+iy&1-z
\end{bmatrix},
$$
where $x :=\text{tr}(\rho\hat{S}_x)$, $y:=\text{tr}(\rho\hat{S}_y)$, and $z:=\text{tr}(\rho\hat{S}_z)$ are the Bloch vector coordinates satisfying $x^2+y^2+z^2\leq 1$. Equivalently, one can consider any system of coordinates induced by the expectation values of three mutually orthogonal spin operators. In our case, one can take $\hat{S}_1$, $\hat{S}_2$, and $\hat{S}_3 := \hat{S}\cdot(\hat{n}_1\times\hat{n}_2)$, 
so that 
\begin{align}
s_1^2+s_2^2+s_3^2 \leq 1,\label{eq:sphere}
\end{align}
where $s_1 := \text{tr}(\rho\hat{S}_1)$, $s_2 := \text{tr}(\rho\hat{S}_2)$, and $s_3 := \text{tr}(\rho\hat{S}_3)$ are the expected values for the spin components of the particle along the directions $\hat{n}_1$, $\hat{n}_2$, and $\hat{n}_1\times\hat{n}_2$, respectively.
\subsection{Wigner distribution $p_{\text{W}}$}
We recall some properties of the Wigner distribution $p_{\text{W}}$ associated with the observables $\hat{S}_1$ and $\hat{S}_2$ and the quantum state $\rho$ studied in Refs. \cite{Cohen1986-COHJWD,10.1063/1.5140632}.\\\\
\textbf{Proposition $8$:} The Wigner distribution $p_{\text{W}}$ associated with the observables $\hat{S}_1$ and $\hat{S}_2$ and the quantum state $\rho$ is formally given by 
\begin{align*}
    p_{\text{W}}(x_1,x_2) = (1+s_1x_1+s_2x_2)p_{\text{W}}^{0}(x_1,x_2), 
\end{align*}
 where
 \begin{align*}
 p_{\text{W}}^{0}(x_1,x_2)
 &=\cfrac{1}{(2\pi)^2}\mathcal{F}(\cos(\|\cdot\|))(x_1,x_2),
 \end{align*}
is the Wigner distribution for any state with Bloch vector normal to the plane defined by $\hat{n}_1$ and $\hat{n}_2$, and  
 \begin{align*}
     \mathcal{F}(\cos(\|\cdot\|))(x_1,x_2) = \int_{\mathbb{R}^2}\cos(\|\xi\|)e^{-i(x_1\xi_1+x_2\xi_2)}\text{d}\xi_1\text{d}\xi_2.
 \end{align*}\textbf{Proof:} By definition, the Wigner distribution $p_{\text{W}}$ is given by 
\begin{align*}
    p_{\text{W}}(x_1,x_2) =\cfrac{1}{(2\pi)^2}\int_{\mathbb{R}^2}f_{\text{W}}(\xi_1,\xi_2)e^{-i(x_1\xi_1+x_2\xi_2)}\text{d}\xi_1\text{d}\xi_2,
\end{align*}
where $f_{\text{W}}$ is given by 
\begin{align*}
f_{\text{W}}(\xi_1,\xi_2) &= \text{tr}(\rho e^{i(\xi_1\hat{S}_1+\xi_2\hat{S}_2)})\\
&= \cos(\|\xi\|) + i(s_1\xi_1+s_2\xi_2)\frac{\sin(\|\xi\|)}{\|\xi\|}\\
&= \left(1-is_1\cfrac{\partial}{\partial \xi_1}-is_2\cfrac{\partial}{\partial \xi_2}\right)\cos(\|\xi\|).
\end{align*}
Performing integration by parts with respect to $\xi_1$ and $\xi_2$ yields  
\begin{align*}
    p_{\text{W}}(x_1,x_2) 
    &=\cfrac{(1+s_1x_1+s_2x_2)}{(2\pi)^2}\mathcal{F}(\cos(\|\cdot\|))(x_1,x_2).
\end{align*}
If a state $\rho^{0}$ corresponds to a Bloch vector normal to the plane defined by  $\hat{n}_1$ and $\hat{n}_2$, then the expectations $s_1^{0}=\text{tr}(\rho^{0}\hat{S}_1)$ and  $s_2^{0}=\text{tr}(\rho^{0}\hat{S}_2)$ are zero, so that
\begin{align*}
p_{\text{W}}^{0}(x_1,x_2) &= \cfrac{1}{(2\pi)^2}\int_{\mathbb{R}^2}\cos(\|\xi\|)e^{-i(x_1\xi_1+x_2\xi_2)}\text{d}\xi_1\text{d}\xi_2\\
&=\cfrac{1}{(2\pi)^2}\mathcal{F}(\cos(\|\cdot\|))(x_1,x_2),
\end{align*}
which yields the desired result.\hfill$\blacksquare$.\\\\ 
Next, using the Paley-Wiener-Schwarz theorem, we prove that $p_{\text{W}}$ is always supported in the unit disk. \\\\
\textbf{Proposition $9$:} The Wigner quasi-probability distribution $p_{\text{W}}$ associated with the observables $\hat{S}_1$ and $\hat{S}_2$ and the quantum state $\rho$ satisfies 
$$
\text{supp}(p_{\text{W}}) \subseteq \mathbb{D},
$$
where $\mathbb{D}\subseteq\mathbb{R}^2$ is the closed unit disk. \\\\
\textbf{Proof:} Let $\|A\|_{\text{F}}:=\sqrt{\text{tr}(AA^{\dag})}$ denote the Frobenius norm. By the Cauchy-Schwarz inequality, $|\text{tr}(AB)|\leq \|A\|_{\text{F}}\|B\|_{\text{F}}$. Then, the function 
$$
f_{\text{W}}(z_1,z_2) = \text{tr}\left(\rho e^{i(z_1\hat{S}_1+z_2\hat{S}_2)}\right),~~z:=(z_1,z_2)\in\mathbb{C}^2,
$$
which is analytic everywhere in $\mathbb{C}^2$, satisfies the estimate 
\begin{align*}
&|f_{\text{W}}(z_1,z_2)|\leq \|\rho\|_{\text{F}}\left\|e^{i(z_1\hat{S}_1+z_2\hat{S}_2)}\right\|_{\text{F}}\\
&= \|\rho\|_{\text{F}}\left\|e^{-(\Im(z_1)\hat{S}_1+\Im(z_2)\hat{S}_2)}\right\|_{\text{F}}\\
&= \|\rho\|_{\text{F}}\left\|\sum_{k=0}^{\infty}\frac{(-1)^k(\Im(z_1)\hat{S}_1+\Im(z_2)\hat{S}_2)^{k}}{k!}\right\|_{\text{F}}.
\end{align*}
By noting that $(\Im(z_1)\hat{S}_1+\Im(z_2)\hat{S}_2)^2 = \|\Im(z)\|^2_2\hat{I}$, we can split the above sum into even and odd parts to get 
\begin{align*}
    &\left\|\sum_{k=0}^{\infty}\frac{(-1)^k(\Im(z_1)\hat{S}_1+\Im(z_2)\hat{S}_2)^{k}}{k!}\right\|_{\text{F}}\hspace{-2mm}\leq \sum_{k=0}^{\infty}\frac{\|\Im(z)\|_2^{2k}}{(2k)!}\|\hat{I}\|_{\text{F}}\\
&+\left\|\Im(z_1)\hat{S}_1\hspace{-1mm}+\hspace{-1mm}\Im(z_2)\hat{S}_2\right\|_{\text{F}}\sum_{k=0}^{\infty}\hspace{-1mm}\frac{\|\Im(z)\|_2^{2k}}{(2k+1)!} \hspace{-1mm}\leq \sqrt{2}\sum_{k=0}^{\infty}\frac{\|\Im(z)\|_2^{k}}{k!}.
\end{align*}
Thus, we obtain the estimate 
\begin{align*}
    |f_{\text{W}}(z_1,z_2)|&\leq\sqrt{2}\|\rho\|_{\text{F}}\sum_{k=0}^{\infty}\frac{\|\Im(z)\|_2^{k}}{k!}\\
&=\sqrt{2}\|\rho\|_{\text{F}}e^{\|\Im(z)\|_2} = \sqrt{2}\|\rho\|_{\text{F}}e^{H_{\mathbb{D}}(\Im(z))}.
\end{align*}
By the Paley-Wiener-Schwarz theorem, $\text{supp}(p_{\text{W}})\subseteq \mathbb{D}$ as desired.\hfill$\blacksquare$\\\\
\noindent\textbf{Remark:} Since $p_{\text{W}}^{0}$ is radially symmetric, the Wigner distribution $p_{\text{W}}$ can be expressed in polar coordinates via the change of variables $x_1=r\cos(\theta)$ and $x_2 = r\sin(\theta)$, where $r = \|x\|$, as
\begin{align*}
    p_{\text{W}}(r,\theta) &= (1+s_1r\cos(\theta)+s_2r\sin(\theta))p_{\text{W}}^{0}(r).
\end{align*}
Thus, the Wigner distribution for any state $\rho$ can always be written as the Wigner distribution $p_{\text{W}}^{0}$ corresponding to, for instance, the maximally mixed state $\rho=I/2$, multiplied by the function
$$
h(r,\theta) = 1+s_1r\cos(\theta)+s_2r\sin(\theta).
$$
 Compare, for example, the two Wigner distributions for the observables $\hat{S}_1 = \hat{S}_x$ and $\hat{S}_2=\hat{S}_y$ shown in {\color{black}the two right subplots of }Fig.~\ref{fig:CohenScully}. Modulo the Gaussian regularization, the bottom Wigner distribution can be obtained by multiplying the top Wigner distribution by
$$
h(r,\theta) = 1+\cfrac{r}{\sqrt{2}}(\cos(\theta)+\sin(\theta)) = 1+r\sin(\theta+\frac{\pi}{4}).
$$
That is because the Wigner distribution for the state 
$$
\rho = \cfrac{1}{2}\begin{bmatrix}
    1& \frac{1-i}{\sqrt{2}}\\
    \frac{1+i}{\sqrt{2}}& 1
\end{bmatrix},
$$
corresponds to the Bloch vector coordinates $x = 1/\sqrt{2}$, $y = 1/\sqrt{2}$, and $z = 0$. Thus, $s_1=s_2=1/\sqrt{2}$.

Lastly, notice that the function $h$ is always non-negative in $\mathbb{D}$ where $p_{\text{W}}$ is supported. To see this, apply the reverse triangle inequality, followed by the Cauchy-Schwarz inequality, to get
\begin{align*}
h(r,\theta)&\geq 1-|s_1r\cos(\theta)+s_2r\sin(\theta)|\\
&\geq 1-|r|\sqrt{s_1^2+s_2^2}\geq 0,
\end{align*}
where the last inequality follows from Eq. \eqref{eq:sphere}. This implies that the sign of the Wigner distribution for $\hat S_1$ and $\hat S_2$, which reflects the quantum nature of the representation, is independent of the state $\rho$.

\subsection{Margenau-Hill quasi-characteristic functions $f_{\text{MH}_m}$ and the Mehler-Heine theorem}
In what follows, we compute closed-form expressions for the Margenau-Hill quasi-characteristic functions $f_{\text{MH}_m}$ and show that their convergence to $f_{\text{W}}$, which is uniform on compact subsets of $\mathbb{C}^2$ by Lemma $1$, is a special case of the Mehler-Heine theorem. We start by proving the following result.\\\\
\textbf{Lemma $2$:}
  Let $m\in\mathbb{N}$ and $\xi_1,\xi_2\in\mathbb{R}$.
  Then,
  \begin{align}
    &\frac{\left(e^{i\frac{\xi_1}{m}\hat{S}_1}e^{i\frac{\xi_2}{m}\hat{S}_2}\right)^m+\left(e^{i\frac{\xi_2}{m}\hat{S}_2}e^{i\frac{\xi_1}{m}\hat{S}_1}\right)^m}{2}\nonumber \\
      &= T_m(a_m) + ib_mU_{m-1}(a_m),\label{eq:cheby}
      \end{align}
      and 
      \begin{align}
      &\frac{\left(e^{i\frac{\xi_1}{m}\hat{S}_1}e^{i\frac{\xi_2}{m}\hat{S}_2}\right)^m-\left(e^{i\frac{\xi_2}{m}\hat{S}_2}e^{i\frac{\xi_1}{m}\hat{S}_1}\right)^m}{2}\nonumber\\
      &= \label{eq:cheby2} U_{m-1}(a_m)\sin\left(\frac{\xi_1}{m}\right)\sin\left(\frac{\xi_2}{m}\right)\hat{S}_2\hat{S}_1,
  \end{align}
  where
  \begin{align*}
      a_m &:= \cos\left(\frac{\xi_1}{m}\right)\cos\left(\frac{\xi_2}{m}\right)\hat{I},\\
      b_m &:=\sin\left(\frac{\xi_1}{m}\right)\cos\left(\frac{\xi_2}{m}\right)\hat{S}_1+\cos\left(\frac{\xi_1}{m}\right)\sin\left(\frac{\xi_2}{m}\right)\hat{S}_2,
  \end{align*}
and $T_m(\cdot)$ and $U_m(\cdot)$ are the $m^{\text{th}}$ degree Chebyshev polynomials of the first and second kinds, respectively.\\\\
  \textbf{Proof:} The proof is given in Appendix \ref{asec:4}.\hfill$\blacksquare$\\
  
  The resemblance of Eq. \eqref{eq:cheby} with the $m^{\text{th}}$ power
  \begin{align*}
      z^m=T_m(a)+ibU_{m-1}(a),
  \end{align*}
 of a complex number $z=a+ib$ of modulus $1$, $a,b\in\mathbb R$, is inescapable. Then, Eqs. \ref{eq:cheby} and \ref{eq:cheby2} together may be seen as a non-commutative version of de Moivre's formula.
 Using the above lemma, we can derive closed-form expressions for the Margenau-Hill quasi-characteristic functions $f_{\text{MH}_m}$ for all $m\in\mathbb{N}$, as explained next.\\\newline 
 \textbf{Proposition $10$:} The Margenau-Hill quasi-characteristic functions $ f_{\text{MH}_m}(\xi_1,\xi_2)$ associated with the spin operators $\hat{S}_1$, $\hat{S}_2$ and the state $\rho$ are given by 
 $$
 \left(1-is_1\frac{\partial}{\partial\xi_1}-is_2\frac{\partial}{\partial\xi_2}\right)T_{m}\left(\cos\left(\frac{\xi_1}{m}\right)\cos\left(\frac{\xi_2}{m}\right)\right),
 $$
   where $T_m(\cdot)$ is the $m^{\text{th}}$ degree Chebyshev polynomial of the first kind, $m\in\mathbb{N}$.\\\\
   \textbf{Proof:} Equation \eqref{eq:cheby} implies that  for all $m\in\mathbb{N}$, 
   \begin{align*}
       &f_{\text{MH}_m}(\xi_1,\xi_2) = \text{tr}(\rho(T_{m}(a_m)+ib_mU_{m-1}(a_m))).
   \end{align*}
Since $\frac{\text{d}}{\text{d}x}T_m(x) = mU_{m-1}(x)$, the right-hand side equals
   \begin{align*}
  &\text{tr}\left(\rho\left(\hat{I}-i\hat{S}_1\frac{\partial}{\partial\xi_1}-i\hat{S}_2\frac{\partial}{\partial\xi_2}\right)T_m\left(\cos\left(\frac{\xi_1}{m}\right)\cos\left(\frac{\xi_2}{m}\hspace{-1mm}\right)\hspace{-1mm}\right)\hspace{-1mm}\right)\\
  &=\left(1-is_1\frac{\partial}{\partial\xi_1}-is_2\frac{\partial}{\partial\xi_2}\right)T_{m}\left(\cos\left(\frac{\xi_1}{m}\right)\cos\left(\frac{\xi_2}{m}\right)\right),
   \end{align*}
   which completes the proof.\hfill$\blacksquare$\\
   
   \noindent{\bf Remark:} So far, we have seen that both the Wigner quasi-characteristic function $f_{\text{W}}$ as well as the Margenau-Hill quasi-characteristic functions $f_{\text{MH}_m}$  can be derived by applying the operator 
   $$
1-is_1\cfrac{\partial}{\partial \xi_1}-is_2\cfrac{\partial}{\partial \xi_2}
   $$
   to the quasi-characteristic function corresponding to any state $\rho_0$ satisfying $\text{tr}(\rho^0\hat{S}_1)= \text{tr}(\rho^0\hat{S}_2)=0$. 
  Thus, knowledge of the quasi-characteristic function for any such $\rho_0$ is enough to construct the quasi-characteristic function for any other state $\rho$.\\
  
  In light of Proposition $10$, we conclude by noting the connection between the Mehler-Heine theorem and the convergence property of the Margenau-Hill quasi-characteristic functions $f_{\text{MH}_m}$ to $f_{\text{W}}$.\\\\
  \textbf{Proposition 11:} For the special case of spin observables $\hat{S}_1$ and $\hat{S}_2$, Lemma $1$ follows directly from the Mehler-Heine theorem.\\\\
  \textbf{Proof:} The proof is given in Appendix \ref{asec:5}.\hfill$\blacksquare$\\\\
  It is unclear whether this connection between the Lie-Trotter product formula and the Mehler-Heine theorem for spin$-1/2$ observables is simply a mathematical coincidence or the manifestation of a deeper fact.

\subsection{Margenau-Hill quasi-probability distributions $p_{\text{MH}_m}$}
We now derive closed-form expressions for the Margenau-Hill quasi-probability distributions $p_{\text{MH}_m}$ for all $m\in\mathbb{N}$. This is done by taking the Fourier transform of the corresponding expressions for $f_{\text{MH}_m}$ in Proposition~$10$.\\\\
\textbf{Proposition $12$:} 
The Margenau-Hill quasi-probability distribution of order $m\in\mathbb N$ associated with the spin observables $\hat S_1,\;\hat S_2$ and
a quantum state $\rho$ is given by
\begin{align*}
    p_{\text{MH}_m} =(1+s_1x_1+s_2x_2)p^0_{\text{MH}_m},
\end{align*}
where $s_i=\text{tr}(\rho\hat S_i)$, $i\in\{1,2\}$,
   \begin{align*}
    p^0_{\text{MH}_m} \hspace{-1mm}=\sum_{n = 0}^m\cfrac{a_{mn}}{4^n}\hspace{-1mm}\left[\sum_{p,q=0}^n\hspace{-1mm}\binom{n}{p}\binom{n}{q}\delta_{\left(x_1-\frac{n-2p}{m},~x_2-\frac{n-2q}{m}\right)}\right],
\end{align*}
and $a_{mn}$ are the coefficients of the Chebyshev polynomial of the first kind of order $m$, that is,
\begin{align*}
    T_m(x) = \sum_{n = 0}^{m}a_{mn} x^n.
\end{align*} 
\textbf{Proof:} By Proposition $10$, the Margenau-Hill quasi-characteristic function $ f_{\text{MH}_m}(\xi_1,\xi_2)$ is given by 
 $$
 \left(1-is_1\frac{\partial}{\partial\xi_1}-is_2\frac{\partial}{\partial\xi_2}\right)T_{m}\left(\cos\left(\frac{\xi_1}{m}\right)\cos\left(\frac{\xi_2}{m}\right)\right),
 $$
for all $m\in\mathbb{N}$. It follows that
\begin{align*}
    p_{\text{MH}_m}&=\frac{1}{(2\pi)^2}\mathcal{F}(f_{\text{MH}_m})=(1+s_1x_1+s_2x_2)p^0_{\text{MH}_m},
    \end{align*}
where
    \begin{align*}
 &p^0_{\text{MH}_m}(x_1,x_2)= \sum_{n = 0}^m\cfrac{a_{mn}}{(2\pi)^2}\mathcal{F}\left(\cos^n\left(\frac{x_1}{m}\right)\cos^n\left(\frac{x_2}{m}\right)\right),\\
    &=\sum_{n = 0}^m\cfrac{a_{mn}}{4^n}\left[\sum_{p,q=0}^n\binom{n}{p}\binom{n}{q}\delta_{\left(x_1-\frac{n-2p}{m},~x_2-\frac{n-2q}{m}\right)}\right].
\end{align*}
The last equality follows from the fact that
$$
\mathcal{F}\left(\cos^n\left(\frac{\cdot}{m}\right)\right) = \cfrac{2\pi}{2^n}\left[\sum_{p=0}^n\binom{n}{p}\delta\left(\cdot-\frac{n-2p}{m}\right)\right],
$$
which completes the proof.\hfill$\blacksquare$
\section{repeated experiments}
In this section, we discuss the Wigner distribution $p_{\text{W}}$ and its particle approximations $p_{\text{MH}_m}$ in the context of repeated experiments.

Recall from Proposition $3$ that the Margenau-Hill quasi-probability distribution $p_{\text{MH}_m}$ of order $m$ is supported on the grid of points 
\begin{align*}
    \cfrac{1}{m}\sum_{i=1}^m\Lambda = \cfrac{\Lambda +\Lambda +\hdots+\Lambda}{m},
\end{align*}
where the summation is in the sense of Minkowski, and
$$
\Lambda := \sigma(\hat{A}_1)\times\sigma(\hat{A}_2)\times\hdots\times\sigma(\hat{A}_n),
$$
with $\sigma(\hat{A}_k)$ the spectrum of $\hat{A}_k$, $k=1,\hdots,n$. The set $\Lambda$ consists of all tuples of eigenvalues and is the support that is expected of a joint probability law on the classical measurement outcomes of the observables $\hat{A}_1,\hdots,\hat{A}_n$. For instance, in the case of spin$-1/2$ observables $\hat{S}_1$ and $\hat{S}_2$  discussed in Sec. \ref{sec:4}, the set $\Lambda$ is given by 
\begin{align*}
\Lambda &= (-1,+1)\times(-1,+1),\\
&= \{(-1,-1),(-1,+1),(+1,-1),(+1,+1)\},\\
&=\{(\pm 1,\pm 1),(\pm 1, \mp 1)\},
\end{align*}
which is the support for the Margenau-Hill quasi-probability distribution $p_{\text{MH}_m}$ of order $m=1$.  

Since the probability measures $p_{\text{MH}_m}$ are sign-indefinite, the underlying experiments are not realizable, and will hence be referred to as \textit{thought experiments}. For instance, the thought experiment associated with the signed measure $p_{\text{MH}_1}$ is the \textit{simultaneous} measurement of the spin components of a spin$-1/2$ particle in state $\rho$ along the directions $\hat{n}_1$ and $\hat{n}_2$, see Ref.~\cite{PhysRevA.49.1562}. 
 
Likewise, the support for $p_{\text{MH}_2}$, which is
\begin{align*}
\cfrac{\Lambda+\Lambda}{2} &= \cfrac{((-1,+1)\times(-1,+1)) + ((-1,+1)\times(-1,+1))}{2},\\
&= \{(\pm 1,\pm 1),(\pm 1, \mp 1),(\pm 1,0),(0,\pm 1),(0,0)\},
\end{align*}
suggests that the associated thought experiment for the case $m=2$ would be the \textit{instantaneous repetition of the thought experiment in $m=1$ twice}, for the same $\rho$, with the average of the two results recorded as the outcome. The simultaneous measurements performed in these thought experiments are not quantum measurements because $\hat{S}_1$ and $\hat{S}_2$ do not commute. As a consequence, the axiom of repetition need not apply, and nine possible outcomes are present. For instance, the outcome $(0,+1)$ may arise as a result of $(+1,+1)$ instantiating in the first simultaneous measurement and $(-1,+1)$ in the second. 

  In a similar manner, the support of $p_{\text{MH}_N}$ for any $N\in\mathbb{N}$ becomes the set of outcomes obtained by \textit{instantaneously repeating the thought experiment for $m=1$, $N$ times, and recording the average.} Thus, the Wigner quasi-probability distribution $p_{\text{W}}$ corresponds to this limiting thought experiment that involves an \textit{infinite} instantaneous repetition of the thought experiment in $m=1$.

Since this limit process averages out the results from theoretically sampling the system in state $\rho$ infinitely many times, the support of the Wigner distribution is still confined in the unit square. See, for instance, Fig.~\ref{fig:resspin9}, which displays the evolution of the support of the Margenau-Hill distributions $p_{\text{MH}_N}$ for $\hat{S}_1$ and $\hat{S}_2$ as $N\rightarrow\infty$. The fact, however, that the support is always inside the \textit{unit Disk}, which is the joint numerical range of $\hat{S}_1$ and $\hat{S}_2$, suggests that with every run of this thought experiment, the resulting average must correspond to the spin components of the particle in some state $\sigma$, as  
$$
\text{tr}\left(\sigma\hat{S}_1\right)^2+\text{tr}\left(\sigma\hat{S}_2\right)^2\leq 1.
$$
We believe that a time-resolved version of this thought experiment could be linked to the continuous monitoring of non-commuting observables \cite{jacobs2006straightforward}. Such an interpretation could provide insight on certain features of $p_\text{W}$, such as its regions of positivity as well as its shape, from a physical perspective. The above discussion is no different for a general tuple of operators $\hat{A}_1,\hdots,\hat{A}_n$, and potential links to the theory of continuous measurement are of great interest. 

Finally, when all of the observables $\hat{A}_1,\hdots,\hat{A}_n$ commute, all successive supports collapse to that of $p_{\text{MH}_1}$ and the sequence of thought experiments, which are now realizable, must lead to outcomes confined to the classical sample space $\Lambda$. In the context of our proposed thought experiments, this implies that making multiple repetitions of the same measurement does not alter the average when the observables commute, i.e., subsequent measurements are identical to the outcomes obtained in the first measurement. This is consistent with the axiom of repetition, which asserts that performing the same measurement on a quantum system will not change the resulting outcome.
 
\section{Conclusion}
In this work, a class of real-valued signed discrete probability measures given by 
\begin{align*}
p_{\text{MH}_{m}} = \cfrac{1}{(2\pi)^n}\mathcal{F}(f_{\text{MH}_m}),
\end{align*}
for $n$ arbitrary quantum observables is derived and
studied based on quasi-characteristic functions $f_{\text{MH}_m}$ with symmetrized operator orderings of Margenau-Hill type. These measures are given by affine combinations of Dirac delta distributions supported over the finite spectral range of the quantum observables, and give the correct probability marginals when coarse-grained along any principal axis. We showed that these particle approximations converge weakly to their corresponding Wigner distribution, and the convergence can be upgraded if they are smeared with an appropriate Schwarz function. Closed-form expressions in the case of bivariate quasi-probability distributions for the spin measurements of spin$-1/2$ particles are provided. As a side result, the convergence of the approximants in this case follows from the Mehler-Heine theorem. 
Finally, we discussed the Wigner distribution and its particle approximations in the context of repeated thought experiments. Namely, the supports of $p_{\text{W}}$ and $p_{\text{MH}_m}$ point towards thought experiments involving repeated simultaneous measurements on the state $\rho$. When the operators mutually commute, these supports reduce to the classical grid of eigenvalues, in agreement with the axiom of repetition. 
\begin{acknowledgments}
This research has been supported in part by the NSF under ECCS-2347357, AFOSR under FA9550-24-1-0278, and ARO under W911NF-22-1-0292.
\end{acknowledgments}

{\onecolumngrid
\appendix 

\section{Proof of Proposition $2$}\label{asec:1}

\noindent We provide herein details for the inequality 
 \begin{align*}
      \sum_{\pi\in S_n}\text{tr}\left(\left|\rho\left(\prod_{k=1}^ne^{i\frac{-z_{\pi(k)}}{m}\hat{A}_{\pi(k)}}\right)^m\right|\right) \leq e^{\max_{v\in \mathcal{V}(K)}\langle v,\Im(z)\rangle}\sum_{\pi\in S_n}\text{tr}\left(\rho\right).
    \end{align*}
   By definition, 
\begin{align*}
    \left|\rho\left(\prod_{k=1}^ne^{i\frac{-z_{\pi(k)}}{m}\hat{A}_{\pi(k)}}\right)^m\right| &= \sqrt{\rho\left(\prod_{k=1}^ne^{i\frac{-z_{\pi(k)}}{m}\hat{A}_{\pi(k)}}\right)^m\left(\prod_{k=1}^ne^{i\frac{\bar{z}_{\pi(n-k+1)}}{m}\hat{A}_{\pi(n-k+1)}}\right)^m\rho}.
\end{align*}
Notice that the innermost pair of factors simplify and are bounded as follows
\begin{align*}
    &\left(\prod_{k=1}^ne^{i\frac{-z_{\pi(k)}}{m}\hat{A}_{\pi(k)}}\right)\left(\prod_{k=1}^ne^{i\frac{\bar{z}_{\pi(n-k+1)}}{m}\hat{A}_{\pi(n-k+1)}}\right) \\&=\left(\prod_{k=1}^{n-1}e^{i\frac{-z_{\pi(k)}}{m}\hat{A}_{\pi(k)}}\right)e^{-2\frac{\Im(z_{\pi(n)})}{m}\hat{A}_{\pi(n)}}\left(\prod_{k=2}^ne^{i\frac{\bar{z}_{\pi(n-k+1)}}{m}\hat{A}_{\pi(n-k+1)}}\right)\\
    &\leq \left(\prod_{k=1}^{n-1}e^{i\frac{-z_{\pi(k)}}{m}\hat{A}_{\pi(k)}}\right)\left(\prod_{k=2}^ne^{i\frac{\bar{z}_{\pi(n-k+1)}}{m}\hat{A}_{\pi(n-k+1)}}\right)e^{2\frac{v_{n}\Im(z_{\pi(n)})}{m}\hat{I}},
\end{align*}
where the last inequality follows because the positive operator  $e^{-2\frac{\Im(z_{\pi(n)})}{m}\hat{A}_{\pi(n)}}$ is less than $e^{2\frac{v_n\Im(z_{\pi(n)})}{m} \hat{I}}$ for some $v_n\in\{\lambda_{\text{min}}(\hat{A}_{\pi(n)}),~\lambda_{\text{max}}(\hat{A}_{\pi(n)})\}$, and applying a congruence transformation will not change this fact; $e^{2\frac{v_n\Im(z_{\pi(n)})}{m} \hat{I}}$ is a scalar multiple of the identity and can be moved to the right.  Repeating this process for the subsequent pairs of innermost factors, we obtain the final estimate 
\begin{align*}
    \left(\prod_{k=1}^ne^{i\frac{-z_{\pi(k)}}{m}\hat{A}_{\pi(k)}}\right)\left(\prod_{k=1}^ne^{i\frac{\bar{z}_{\pi(n-k+1)}}{m}\hat{A}_{\pi(n-k+1)}}\right) \leq e^{2\frac{\max_{v\in \mathcal{V}(K)}\langle v,\Im(z)\rangle} {m}\hat{I}}.
\end{align*}
Thus,
\begin{align*}
    \left(\prod_{k=1}^ne^{i\frac{-z_{\pi(k)}}{m}\hat{A}_{\pi(k)}}\right)^m\left(\prod_{k=1}^ne^{i\frac{\bar{z}_{\pi(n-k+1)}}{m}\hat{A}_{\pi(n-k+1)}}\right)^m \leq e^{2\max_{v\in \mathcal{V}(K)}\langle v,\Im(z)\rangle\hat{I}}.
\end{align*}
Multiplying from the left and right by $\rho$ and recalling that the $\sqrt{\cdot}$ function is operator monotone yields 
\begin{align*}
      \left|\rho\left(\prod_{k=1}^ne^{i\frac{-z_{\pi(k)}}{m}\hat{A}_{\pi(k)}}\right)^m\right| \leq \sqrt{\rho e^{2\max_{v\in \mathcal{V}(K)}\langle v,\Im(z)\rangle\hat{I}}\rho} = e^{\max_{v\in \mathcal{V}(K)}\langle v,\Im(z)\rangle}\rho.
\end{align*}
Taking the trace on both sides and summing over all $\pi\in S_n$ yields the desired inequality.
\section{Proof of Lemma $1$}\label{asec:2}
\noindent Let $K$ be any compact set in $\mathbb{C}^n$ and let $\|\cdot\|_{\text{F}}$ denote the Frobenius norm
$$
\|A\|_{\text{F}} := \sqrt{\text{tr}(AA^{\dag})},~A\in\mathbb{C}^{d\times d}.
$$
Recall that $|\text{tr}(AB)|\leq \|A\|_{\text{F}}\|B\|_{\text{F}}$ and that the functions $f_{\text{W}}$ and $f_{\text{MH}_m}$ can be analytically extended to $\mathbb{C}^n$ for all $m\in\mathbb{N}$. Then, 
\begin{align*}
    &\sup_{\xi\in K}|f_{\text{W}}(\xi)-f_{\text{MH}_m}(\xi)|= \sup_{\xi\in K}\left|\text{tr}\left(\rho e^{i\xi\cdot {\hat{A}}}\right)-\cfrac{1}{n!}~\text{tr}\left(\rho\sum_{\pi \in S_n}\left(\prod_{k=1}^ne^{i\frac{\xi_{\pi(k)}}{m}\hat{A}_{\pi(k)}}\right)^m\right)\right|\\
    &\leq \cfrac{1}{n!}\sum_{\pi\in S_n}\sup_{\xi\in K}\left|\text{tr}\left(\rho\left(e^{i\xi\cdot {\hat{A}}}-\left(\prod_{k=1}^ne^{i\frac{\xi_{\pi(k)}}{m}\hat{A}_{\pi(k)}}\right)^m\right)\right)\right|\leq \cfrac{\|\rho\|_{\text{F}}}{n!}\sum_{\pi\in S_n}\sup_{\xi\in K}\left\|e^{i\xi\cdot {\hat{A}}}-\left(\prod_{k=1}^ne^{i\frac{\xi_{\pi(k)}}{m}\hat{A}_{\pi(k)}}\right)^m\right\|_{\text{F}}.
\end{align*}
Thus, it is enough to show that 
$$
\sup_{\xi\in K}\left\|e^{i\xi\cdot {\hat{A}}}-\left(\prod_{k=1}^ne^{i\frac{\xi_{\pi(k)}}{m}\hat{A}_{\pi(k)}}\right)^m\right\|_{\text{F}} \rightarrow 0~~\text{as}~~m\rightarrow\infty,
$$
for all permutations 
$\pi\in S_n$. By symmetry, it is sufficient to consider the trivial permutation only. To that end, define 
$$
C := e^{i\frac{\xi}{m}\cdot {\hat{A}}}~~\text{and}~~D := 
\prod_{k=1}^ne^{i\frac{\xi_{k}}{m}\hat{A}_k}.
$$
Then, applying the Cauchy product formula on $D$ yields
$$ 
D = \prod_{k=1}^n\left(\sum_{j_k=0}^{\infty}\cfrac{(i\frac{\xi_k}{m}\hat{A}_k)^{j_k}}{j_k!}\right) = \sum_{|j|=0}^{\infty}\left(\prod_{k=1}^n
\cfrac{1}{m^{j_k}}\cfrac{(i\xi_k\hat{A}_k)^{j_k}}{j_k!}\right)
=\sum_{|j|=0}^{\infty}\cfrac{1}{m^{|j|}}\left(\prod_{k=1}^n\cfrac{(i\xi_k\hat{A}_k)^{j_k}}{j_k!}\right),
$$ 
where $|j| := j_1+\cdots+j_n$. Thus, we get
\begin{align*}
 \|C-D\|_{\text{F}} &= \left\|\sum_{|j|=0}^{\infty} \cfrac{1}{m^{|j|}}\frac{(i\xi\cdot\hat{A})^{|j|}}{|j|!} - \sum_{|j|=0}^{\infty}\cfrac{1}{m^{|j|}}\left(\prod_{k=1}^n\cfrac{(i\xi_k\hat{A}_k)^{j_k}}{j_k!}\right)\right\|_{\text{F}} =\left\|\sum_{|j|=2}^{\infty} \cfrac{1}{m^{|j|}}\left(\frac{(i\xi\cdot\hat{A})^{|j|}}{|j|!}-\prod_{k=1}^n\cfrac{(i\xi_k\hat{A}_k)^{j_k}}{j_k!}\right)\right\|_{\text{F}}  \\
 &\leq  \cfrac{1}{m^2}\sum_{|j|=0}^{\infty} \left\|\frac{(i\xi\cdot\hat{A})^{|j|}}{|j|!}-\prod_{k=1}^n\cfrac{(i\xi_k\hat{A}_k)^{j_k}}{j_k!}\right\|_{\text{F}} \leq  \cfrac{1}{m^2}\sum_{|j|=0}^{\infty} \left(\frac{(|\xi|\cdot\|\hat{A}\|_{\text{F}})^{|j|}}{|j|!}+\prod_{k=1}^n\cfrac{(|\xi_k|\|\hat{A}_k\|_{\text{F}})^{j_k}}{j_k!}\right) \\
 &=   \cfrac{1}{m^2}\left(e^{|\xi|\cdot\|\hat{A}\|_{\text{F}}}+\sum_{|j|=0}^{\infty} \left(\prod_{k=1}^n\cfrac{(|\xi_k|\|\hat{A}_k\|_{\text{F}})^{j_k}}{j_k!}\right)\right)= \cfrac{2}{m^2}e^{|\xi|\cdot\|\hat{A}\|_{\text{F}}},
\end{align*}
where $|\xi|\cdot\|\hat{A}\|_{\text{F}}:=|\xi_1|\|\hat{A}_1\|_{\text{F}}+\cdots+|\xi_n|\|\hat{A}_n\|_{\text{F}}$. By noting that $\|C\|_{\text{F}},\|D\|_{\text{F}}\leq 1$, we get 
$$
\| C^{m} - D^{m} \|_{\text{F}} = \left\| \sum_{k=0}^{m-1} C^{k} (C-D) D^{m-k-1}\right\|_{\text{F}} \leq m \|C-D\|_{\text{F}}\leq \cfrac{2}{m}e^{|\xi|\cdot\|\hat{A}\|_{\text{F}}}.$$
Taking the supremum of both
sides over $\xi\in K$ and letting $m\rightarrow\infty$ yields the desired result.\hfill$\blacksquare$
\section{Proof of Proposition $6$}\label{asec:3}
\begin{itemize}
    \item Proof that $\hat{S}_j^2 = \hat{I}$:
    \begin{align*}
    \hat{S}_j^2 = \begin{bmatrix}
        \cos(\theta_j)&\phantom{-}e^{-i\phi_j}\sin(\theta_j)\\
        e^{i\phi_j}\sin(\theta_j)&-\cos(\theta_j)
    \end{bmatrix}\begin{bmatrix}
        \cos(\theta_j)&\phantom{-}e^{-i\phi_j}\sin(\theta_j)\\
        e^{i\phi_j}\sin(\theta_j)&-\cos(\theta_j) \end{bmatrix}= \begin{bmatrix}
        1 &0\\
      0&1 
    \end{bmatrix}= \hat{I}.
\end{align*}
\item Proof that $\text{det}(\hat{S}_j) = -1$:
\begin{align*}
    \text{det}(\hat{S}_j) = \begin{vmatrix}
         \cos(\theta_j)&\phantom{-}e^{-i\phi_j}\sin(\theta_j)\\
        e^{i\phi_j}\sin(\theta_j)&-\cos(\theta_j)
    \end{vmatrix} = -\cos^2(\theta_j) - \sin^2(\theta_j) = -1.
\end{align*}
\item Proof that $\text{tr}(\hat{S}_j) = 0$:
\begin{align*}
    \text{tr}(\hat{S}_j) =\text{tr}\left(\begin{bmatrix}
        \cos(\theta_j)&\phantom{-}e^{-i\phi_j}\sin(\theta_j)\\
        e^{i\phi_j}\sin(\theta_j)&-\cos(\theta_j)
    \end{bmatrix}\right) = \cos(\theta_j)-\cos(\theta_j) = 0.
\end{align*}
\item Proof that $\left[\hat{S}_j,\hat{S}_k\right] = 2i\hat{S}\cdot(\hat{n}_j\times\hat{n}_k)$: If $j=k$, then 
\begin{align*}
    \left[\hat{S}_j,\hat{S}_j\right] = \hat{S}_j^2-\hat{S}_j^2 = 0 = 2i\hat{S}\cdot(\hat{n}_j\times\hat{n}_j).
\end{align*}
If $j\neq k$, recall that the vector $\hat{n}_j\times\hat{n}_k$ is given by 
\begin{align*}
    \hat{n}_j\times\hat{n}_k &= \begin{vmatrix}
\hat{x}&\hat{y}&
\hat{z}\\
\sin(\theta_j)\cos(\phi_j)&\sin(\theta_j)\sin(\phi_j)&\cos(\theta_j)\\
\sin(\theta_k)\cos(\phi_k)&\sin(\theta_k)\sin(\phi_k)&\cos(\theta_k)
    \end{vmatrix}\\
    &=(\sin(\theta_j)\sin(\phi_j)\cos(\theta_k)-\cos(\theta_j)\sin(\theta_k)\sin(\phi_k))\hat{x}-(\sin(\theta_j)\cos(\phi_j)\cos(\theta_k)-\cos(\theta_j)\sin(\theta_k)\cos(\phi_k))\hat{y}\\
    &+(\sin(\phi_k-\phi_j)\sin(\theta_j)\sin(\theta_k))\hat{z}\\
    &:= (\hat{n}_j\times\hat{n}_k)_x\hat{x}+(\hat{n}_j\times\hat{n}_k)_y\hat{y}+(\hat{n}_j\times\hat{n}_k)_z\hat{z}.
\end{align*}
Then, 
\begin{align*}
    &\left[\hat{S}_j,\hat{S}_k\right] = \hat{S}_j\hat{S}_k-\hat{S}_k\hat{S}_j \\
    &= \begin{bmatrix}
        \cos(\theta_j)&\phantom{-}e^{-i\phi_j}\sin(\theta_j)\\
        e^{i\phi_j}\sin(\theta_j)&-\cos(\theta_j)
    \end{bmatrix}\begin{bmatrix}
        \cos(\theta_k)&\phantom{-}e^{-i\phi_k}\sin(\theta_k)\\
        e^{i\phi_k}\sin(\theta_k)&-\cos(\theta_k)
    \end{bmatrix}\\
    &-\begin{bmatrix}
        \cos(\theta_k)&\phantom{-}e^{-i\phi_k}\sin(\theta_k)\\
        e^{i\phi_k}\sin(\theta_k)&-\cos(\theta_k)
    \end{bmatrix}\begin{bmatrix}
        \cos(\theta_j)&\phantom{-}e^{-i\phi_j}\sin(\theta_j)\\
        e^{i\phi_j}\sin(\theta_j)&-\cos(\theta_j)
    \end{bmatrix}\\
    &= \begin{bmatrix}
        \cos(\theta_j)\cos(\theta_k)+e^{i(\phi_k-\phi_j)}\sin(\theta_j)\sin(\theta_k) & e^{-i\phi_k}\cos(\theta_j)\sin(\theta_k)-e^{-i\phi_j}\sin(\theta_j)\cos(\theta_k)\\
        e^{i\phi_j}\sin(\theta_j)\cos(\theta_k)-e^{i\phi_k}\cos(\theta_j)\sin(\theta_k) & \cos(\theta_j)\cos(\theta_k)+e^{i(\phi_j-\phi_k)}\sin(\theta_j)\sin(\theta_k)
    \end{bmatrix}\\
    &- \begin{bmatrix}
        \cos(\theta_j)\cos(\theta_k)+e^{-i(\phi_k-\phi_j)}\sin(\theta_j)\sin(\theta_k) & e^{-i\phi_j}\cos(\theta_k)\sin(\theta_j)-e^{-i\phi_k}\sin(\theta_k)\cos(\theta_j)\\
        e^{i\phi_k}\sin(\theta_k)\cos(\theta_j)-e^{i\phi_j}\cos(\theta_k)\sin(\theta_j) & \cos(\theta_j)\cos(\theta_k)+ e^{-i(\phi_j-\phi_k)}\sin(\theta_j)\sin(\theta_k)
    \end{bmatrix}\\
    &= 2i\begin{bmatrix}
        \sin(\phi_k-\phi_j)\sin(\theta_j)\sin(\theta_k)&ie^{-i\phi_j}\sin(\theta_j)\cos(\theta_k)-ie^{-i\phi_k}\cos(\theta_j)\sin(\theta_k)\\
        -ie^{i\phi_j}\sin(\theta_j)\cos(\theta_k)+ie^{i\phi_k}\cos(\theta_j)\sin(\theta_k)&\sin(\phi_j-\phi_k)\sin(\theta_j)\sin(\theta_k)
    \end{bmatrix}\\
    &= 2i\left(\hat{S}_{\hat{x}}(\hat{n}_j\times\hat{n}_k)_{\hat{x}}+\hat{S}_{\hat{y}}(\hat{n}_j\times\hat{n}_k)_{\hat{y}}+\hat{S}_z(\hat{n}_j\times\hat{n}_k)_{\hat{z}}\right) = 2i\hat{S}\cdot(\hat{n}_j\times\hat{n}_k),
\end{align*}
where 
$$
\hat{S}_{\hat{x}} = \begin{bmatrix}
    0&1\\1&0
\end{bmatrix},~\hat{S}_{\hat{y}} = \begin{bmatrix}
    0&-i\\i&0
\end{bmatrix},~\hat{S}_{\hat{z}} = \begin{bmatrix}
    1&0\\0&-1
\end{bmatrix}.
$$
\item Proof that $\{\hat{S}_j,\hat{S}_k\}=2(\hat{n}_j\cdot\hat{n}_k)\hat{I}$: If $j\neq k$, then 
$$
\{\hat{S}_j,\hat{S}_k\} = \hat{S}_j\hat{S}_k+\hat{S}_k\hat{S}_j =  \hat{S}_j\hat{S}_k-\hat{S}_j\hat{S}_k = 0 = 2(\hat{n}_j\cdot\hat{n}_k)\hat{I}.
$$
If $j = k$, then 
$$
\{\hat{S}_j,\hat{S}_j\} = 2\hat{S}_j^2 =  2\hat{I} = 2(\hat{n}_j\cdot\hat{n}_j)\hat{I}.
$$
\item Proof that $e^{i(\xi_1\hat{S}_1+\xi_2\hat{S}_2)} = \cos(\|\xi\|)\hat{I} + i(\xi_1\hat{S}_1+\xi_2\hat{S}_2)\frac{\sin(\|\xi\|)}{\|\xi\|}$: Since
$$
(\xi_1\hat{S}_1+\xi_2\hat{S}_2)^2 = \xi_1^2\hat{I}+ \xi_1\xi_2\{\hat{S}_1,\hat{S}_2\}+\xi_2^2\hat{I}= \|\xi\|^2\hat{I},
$$Then, 
\begin{align*}
    e^{i(\xi_1\hat{S}_1+\xi_2\hat{S}_2)} &= \sum_{k=0}^{\infty}\cfrac{(i(\xi_1\hat{S}_1+\xi_2\hat{S}_2))^k}{k!} = \sum_{k=0}^{\infty}\cfrac{(i(\xi_1\hat{S}_1+\xi_2\hat{S}_2))^{2k}}{{(2k)}!}+\sum_{k=0}^{\infty}\cfrac{(i(\xi_1\hat{S}_1+\xi_2\hat{S}_2))^{2k+1}}{{(2k+1)}!}\\
    &=\sum_{k=0}^{\infty}\cfrac{(-1)^k\|\xi\|^{2k}\hat{I}}{{(2k)}!}+i(\xi_1\hat{S}_1+\xi_2\hat{S}_2)\sum_{k=0}^{\infty}\cfrac{(-1)^k\|\xi\|^{2k}}{{(2k+1)}!}\\
    &=\sum_{k=0}^{\infty}\cfrac{(-1)^k\|\xi\|^{2k}}{{(2k)}!}\hat{I}+i(\xi_1\hat{S}_1+\xi_2\hat{S}_2)\cfrac{1}{\|\xi\|}\sum_{k=0}^{\infty}\cfrac{(-1)^k\|\xi\|^{2k+1}}{{(2k+1)}!} = \cos(\|\xi\|)\hat{I}+ i(\xi_1\hat{S}_1+\xi_2\hat{S}_2)\cfrac{\sin(\|\xi\|)}{\|\xi\|}.
\end{align*}
 If either $\xi_1$ or $\xi_2$ is set to $0$, we get the familiar identity $e^{i\xi_j\hat{S}_j} = \cos(\xi_j)\hat{I}+ i\hat{S}_j\sin(\xi_j)$,  $j\in\{1,2\}$.
\end{itemize}
\section{Proof of Lemma $2$}\label{asec:4}
\noindent Note first that Eq. \eqref{eq:cheby2} can be re-written as 
  \begin{align}
      &\frac{\left(e^{i\frac{\xi_1}{m}\hat{S}_1}e^{i\frac{\xi_2}{m}\hat{S}_2}\right)^m\hat{S}_1\hat{S}_2+\left(e^{i\frac{\xi_2}{m}\hat{S}_2}e^{i\frac{\xi_1}{m}\hat{S}_1}\right)^m\hat{S}_2\hat{S}_1}{2}\label{eq:cheby3} =U_{m-1}(a_m)\sin\left(\frac{\xi_1}{m}\right)\sin\left(\frac{\xi_2}{m}\right),
  \end{align}
by simply multiplying Eq. \eqref{eq:cheby2} from the right by $\hat{S}_1\hat{S}_2$. We want to show that Eqs. \eqref{eq:cheby} and \eqref{eq:cheby3} hold for all $m\in\mathbb{N}$. To that end, the proof will proceed by induction on $m$. For $m=1$, we have
  \begin{align*}
      a_1 = \cos\left(\xi_1\right)\cos\left(\xi_2\right)\hat I,~~~b_1=\hat{S}_1\sin\left(\xi_1\right)\cos\left(\xi_2\right)+\hat{S}_2\sin\left(\xi_2\right)\cos\left(\xi_1\right).
  \end{align*}
 Then, by recalling that 
 \begin{align*}
     e^{i\xi_1\hat{S}_1} e^{i\xi_2\hat{S}_2} 
     &= \left(\cos\left(\xi_1\right)\hat{I}+i\hat{S}_1\sin\left(\xi_1\right)\right)\left(\cos\left(\xi_2\right)\hat{I}+i\hat{S}_2\sin\left(\xi_2\right)\right)\\
     &= a_{1} + ib_{1} -\sin(\xi_1)\sin(\xi_2)\hat{S}_1\hat{S}_2,
 \end{align*}
 we find that Eqs.~\eqref{eq:cheby} and \eqref{eq:cheby3} hold in this case, namely,
  \begin{align*}
\cfrac{e^{i\xi_1\hat{S}_1}e^{i\xi_2\hat{S}_2}+e^{i\xi_2\hat{S}_2}e^{i\xi_1\hat{S}_1}}{2} &= a_1 + ib_1 =T_1(a_1) + ib_1U_{0}(a_1),\\
\cfrac{e^{i\xi_1\hat{S}_1}e^{i\xi_2\hat{S}_2}\hat{S}_1\hat{S}_2+e^{i\xi_2\hat{S}_2}e^{i\xi_1\hat{S}_1}\hat{S}_2\hat{S}_1}{2} &=  U_{0}(a_1)\sin(\xi_1)\sin(\xi_2).
  \end{align*}
  Next, suppose that Eqs. \eqref{eq:cheby} and \eqref{eq:cheby3} are true for some $m\in\mathbb{N}$, and define  
  \begin{align*}
      \xi_1^{'} := \cfrac{\xi_1 m}{m+1}\text{  ,  }\xi_2^{'} :=\cfrac{\xi_2 m}{m+1}\text{  ,  }s_1:= \sin \left(\cfrac{\xi_1}{m+1}\right)\text{  ,  }s_2:=\sin \left(\cfrac{\xi_2}{m+1}\right).
 \end{align*}
 Then, by recalling that 
 \begin{align*}
     e^{i\frac{\xi^{'}_1}{m}\hat{S}_1} e^{i\frac{\xi_2^{'}}{m}\hat{S}_2} &= \left(\cos\left(\frac{\xi^{'}_1}{m}\right)\hat{I}+i\hat{S}_1\sin\left(\frac{\xi^{'}_1}{m}\right)\right)\left(\cos\left(\frac{\xi^{'}_2}{m}\right)\hat{I}+i\hat{S}_2\sin\left(\frac{\xi^{'}_2}{m}\right)\right)\\
     &= \left(\cos\left(\frac{\xi_1}{m+1}\right)\hat{I}+i\hat{S}_1\sin\left(\frac{\xi_1}{m+1}\right)\right)\left(\cos\left(\frac{\xi_2}{m+1}\right)\hat{I}+i\hat{S}_2\sin\left(\frac{\xi_2}{m+1}\right)\right)\\
     &= a_{m+1} + ib_{m+1} -s_1s_2\hat{S}_1\hat{S}_2,
 \end{align*}
 we find that 
  \begin{align*}
        &\cfrac{\left(e^{i\frac{\xi_1}{m+1}\hat{S}_1}e^{i\frac{\xi_2}{m+1}\hat{S}_2}\right)^{m+1}+\left(e^{i\frac{\xi_2}{m+1}\hat{S}_2}e^{i\frac{\xi_1}{m+1}\hat{S}_1}\right)^{m+1}}{2} = \cfrac{\left(e^{i\frac{\xi_1^{'}}{m}\hat{S}_1}e^{i\frac{\xi_2^{'}}{m}\hat{S}_2}\right)^{m}e^{i\frac{\xi_1^{'}}{m}\hat{S}_1}e^{i\frac{\xi_2^{'}}{m}\hat{S}_2}+\left(e^{i\frac{\xi_2^{'}}{m}\hat{S}_2}e^{i\frac{\xi_1^{'}}{m}\hat{S}_1}\right)^{m}e^{i\frac{\xi_2^{'}}{m}\hat{S}_2}e^{i\frac{\xi_1^{'}}{m}\hat{S}_1}}{2}\\
        &= \underbrace{\cfrac{\left(e^{i\frac{\xi_1^{'}}{m}\hat{S}_1}e^{i\frac{\xi_2^{'}}{m}\hat{S}_2}\right)^{m}+\left(e^{i\frac{\xi_2^{'}}{m}\hat{S}_2}e^{i\frac{\xi_1^{'}}{m}\hat{S}_1}\right)^{m}}{2}}_{\text{use Eq.}\, \eqref{eq:cheby}}\left(a_{m+1}+ib_{m+1}\right)-s_1s_2\underbrace{\cfrac{\left(e^{i\frac{\xi_1^{'}}{m}\hat{S}_1}e^{i\frac{\xi_2^{'}}{m}\hat{S}_2}\right)^{m}\hat{S}_1\hat{S}_2+\left(e^{i\frac{\xi_2^{'}}{n}\hat{S}_2}e^{i\frac{\xi_1^{'}}{m}\hat{S}_1}\right)^{m}\hat{S}_2\hat{S}_1}{2}}_{\text{use Eq.}\, \eqref{eq:cheby3}}\\
           &= (T_{m}(a_{m+1})+ib_{m+1}U_{m-1}(a_{m+1}))\left(a_{m+1}+ib_{m+1}\right)-s_1s_2(U_{m-1}(a_{m+1})s_1s_2)\\
          &= a_{m+1}T_{m}(a_{m+1}) - (b_{m+1}^2+s_1^2s_2^2)U_{m-1}(a_{m+1}) + ib_{m+1}(a_{m+1}U_{m-1}(a_{m+1})+T_{m}(a_{m+1}))\\
        &= a_{m+1}T_{m}(a_{m+1}) - (\hat{I}-(a_{m+1})^2)U_{m-1}(a_{m+1}) + ib_{m+1}(a_{m+1}U_{m-1}(a_{m+1})+T_{m}(a_{m+1}))\\
        &=T_{m+1}(a_{m+1})+ib_{m+1}U_{m}(a_{m+1}),
  \end{align*}
which proves Eq. \eqref{eq:cheby} for $m+1$. The last equality follows from the fact that the Chebyshev polynomials satisfy 
  \begin{align*}
      T_{m+1}(x) = xT_{m}(x) - (1-x^2)U_{m-1}(x)~~~\text{    and   
 }~~~
      U_{m}(x) = xU_{m-1}(x) + T_m(x),~~~\forall x\in\mathbb{R}.
  \end{align*}
 Likewise, we have \begin{align*}
      &\cfrac{\left(e^{i\frac{\xi_1}{m+1}\hat{S}_1}e^{i\frac{\xi_2}{m+1}\hat{S}_2}\right)^{m+1}\hat{S}_1\hat{S}_2+\left(e^{i\frac{\xi_2}{m+1}\hat{S}_2}e^{i\frac{\xi_1}{m+1}\hat{S}_1}\right)^{m+1}\hat{S}_2\hat{S}_1}{2} \\
      &=\cfrac{e^{i\frac{\xi_1^{'}}{m}\hat{S}_1}e^{i\frac{\xi_2^{'}}{m}\hat{S}_2}\left(e^{i\frac{\xi_1^{'}}{m}\hat{S}_1}e^{i\frac{\xi_2^{'}}{m}\hat{S}_2}\right)^{m}\hat{S}_1\hat{S}_2+e^{i\frac{\xi_2^{'}}{m}\hat{S}_2}e^{i\frac{\xi_1^{'}}{m}\hat{S}_1}\left(e^{i\frac{\xi_2^{'}}{m}\hat{S}_2}e^{i\frac{\xi_1^{'}}{m}\hat{S}_1}\right)^{m}\hat{S}_2\hat{S}_1}{2}\\
      &= -s_1s_2\underbrace{\cfrac{\hat{S}_1\hat{S}_2\left(e^{i\frac{\xi_1^{'}}{m}\hat{S}_1}e^{i\frac{\xi_2^{'}}{m}\hat{S}_2}\right)^{m}\hat{S}_1\hat{S}_2+\hat{S}_2\hat{S}_1\left(e^{i\frac{\xi_2^{'}}{m}\hat{S}_2}e^{i\frac{\xi_1^{'}}{m}\hat{S}_1}\right)^{m}\hat{S}_2\hat{S}_1}{2}}_{\text{use Eq. } \eqref{eq:cheby}}\\&+(a_{m+1}+ib_{m+1})\underbrace{\cfrac{\left(e^{i\frac{\xi_1^{'}}{m}\hat{S}_1}e^{i\frac{\xi_2^{'}}{m}\hat{S}_2}\right)^{m}\hat{S}_1\hat{S}_2+\left(e^{i\frac{\xi_2^{'}}{m}\hat{S}_2}e^{i\frac{\xi_1^{'}}{m}\hat{S}_1}\right)^{m}\hat{S}_2\hat{S}_1}{2}}_{\text{use Eq. } \eqref{eq:cheby3}}\\
      &=s_1s_2(T_{m}(a_{m+1})-ib_{m+1}U_{m-1}(a_{m+1})) + (a_{m+1}+ib_{m+1})( U_{m-1}(a_{m+1})s_1s_2)\\
      &= s_1s_2(a_{m+1}U_{m-1}(a_{m+1})+T_{m}(a_{m+1})) =s_1s_2U_{m}(a_{m+1}),
  \end{align*}
  which proves Eq. \eqref{eq:cheby3} for $m+1$. By induction, \eqref{eq:cheby} and \eqref{eq:cheby3} hold for all $m\in \mathbb{N}$.
\section{Alternate Proof for Lemma $1$  when $\hat{A}_1=\hat{S}_1$ and $\hat{A}_2=\hat{S}_2$}\label{asec:5}

Recall that the Mehler-Heine theorem \eqref{eq:Mehler} states that 
\begin{align*}
    \lim_{m\rightarrow \infty}m^{-\alpha}P_m^{(\alpha,\beta)}\left(\cos\left(\cfrac{z}{m}\right)\right)  =\left(\cfrac{z}{2}\right)^{-\alpha}J_{\alpha}(z),
\end{align*}
 uniformly on compact subsets of $\mathbb{C}$. As pointed out in Ref. \cite{STREIT1984393}, G. Szeg\"o's proof of the theorem establishes that 
 \begin{align*}
     \lim_{m\rightarrow \infty}m^{-\alpha}P_m^{(\alpha,\beta)}\left(1-\cfrac{z^2}{2m^2}+o(m^{-2})\right)  =\left(\cfrac{z}{2}\right)^{-\alpha}J_{\alpha}(z).
 \end{align*}
Next, since 
     \begin{align*}
         \cos\left(\frac{z_1}{m}\right)\cos\left(\frac{z_2}{m}\right)=1-\cfrac{z_1^2+z_2^2}{2m^2} +o(m^{-2}),
     \end{align*}
    where the last equality follows by substituting each factor with its Maclaurin series, we get  
    \begin{align}
        \lim_{m\rightarrow \infty}m^{-\alpha}P_m^{(\alpha,\beta)}\left(  \cos\left(\frac{z_1}{m}\right)\cos\left(\frac{z_2}{m}\right)\right) 
 &= \lim_{m\rightarrow \infty}m^{-\alpha}P_m^{(\alpha,\beta)}\left(1-\cfrac{z_1^2+z_2^2}{2m^2} +o(m^{-2})\right)\nonumber\\
 &=   \left(\cfrac{\sqrt{z_1^2+z_2^2}}{2}\right)^{-\alpha}J_{\alpha}\left(\sqrt{z_1^2+z_2^2}\right)\label{eq:Jacob},
     \end{align}
 uniformly on compact subsets of $\mathbb{C}^{2}$. The analogue of \eqref{eq:Jacob} was established in Ref. \cite{STREIT1984393} for a ratio of cosines, instead of a product. Setting $\alpha = \beta = - 1/2$, 
 \begin{align*}
    J_{-1/2}(z) = \sqrt{\cfrac{2}{\pi z}}\cos(z),~~~ P_m^{(-1/2,-1/2)}(z)=\cfrac{(2m)!}{2^{2m}(m!)^2}~T_m(z),
 \end{align*}
 where $T_m(\cdot)$ denotes the $m^{\text{th}}$ degree Chebyshev polynomial of the first kind. Then, Eq. \eqref{eq:Jacob} becomes 
\begin{align}\label{eq:Jacob2}
     \lim_{m\rightarrow \infty}\cfrac{\sqrt{m\pi}(2m)!}{2^{2m}(m!)^2}T_m\left( \cos\left(\frac{z_1}{m}\right)\cos\left(\frac{z_2}{m}\right)\right)=\cos\left(\sqrt{z_1^2+z_2^2}\right).
\end{align}
 Stirling's formula gives that
\begin{align*}
\lim_{m\rightarrow\infty}\cfrac{\sqrt{m\pi}(2m)!}{2^{2m}(m!)^2} = 1,
\end{align*}
and hence, Eq.\ \eqref{eq:Jacob2} reduces to 
$$
\lim_{m\rightarrow\infty}T_{m}\left(\cos\left(\frac{ z_1}{m}\right)\cos\left(\frac{ z_2}{m}\right)\right)= \cos\left(\sqrt{z_1^2+z_2^2}\right),
$$
Since the uniform convergence of a sequence of analytic functions $f_m$ to $f$ on compact subsets of $\mathbb{C}^2$ implies the same type of convergence for their derivatives to $\frac{\partial}{\partial z}f$,  
$$
\lim_{m\rightarrow\infty}\cfrac{\partial}{\partial z_i}T_{m}\left(\cos\left(\frac{ z_1}{m}\right)\cos\left(\frac{ z_2}{m}\right)\right)=\cfrac{\partial}{\partial z_i}\cos\left(\sqrt{z_1^2+z_2^2}\right),
$$
 uniformly on compact subsets of $\mathbb{C}^2$, $i\in\{1,2\}$. Hence,
 \begin{align*}
&\lim_{m\rightarrow\infty}\left(1-is_1\frac{\partial}{\partial z_1}-is_2\frac{\partial}{\partial z_2}\right)T_{m}\left(\cos\left(\frac{ z_1}{m}\right)\cos\left(\frac{ z_2}{m}\right)\right)=\left(1-is_1\frac{\partial}{\partial z_1}-is_2\frac{\partial}{\partial z_2}\right)\cos\left(\sqrt{z_1^2+z_2^2}\right),
 \end{align*}
 uniformly on compact subsets of $\mathbb{C}^2$, which is exactly the statement in Lemma $1$ for the spin operators $\hat{S}_1$ and $\hat{S}_2$.}
 \twocolumngrid
\bibliography{references}
\end{document}